\documentclass[%
 aip,
 amsmath,amssymb,
 reprint,%
]{revtex4-1}

\usepackage{graphicx}
\usepackage{dcolumn}
\usepackage{bm}

\usepackage[utf8]{inputenc}
\usepackage[T1]{fontenc}
\usepackage{mathptmx}
\usepackage{bbm}
\usepackage{amsmath}
\usepackage{amssymb}
\usepackage{mathrsfs}
\usepackage{bm}
\usepackage{bbold}
\usepackage{empheq}
\usepackage{amsfonts}
\usepackage{etoolbox}
\usepackage[colorlinks=true]{hyperref}
\newcommand{\RomanNumeralCaps}[1]{\MakeUppercase{\romannumeral #1}}

\makeatletter
\def\@email#1#2{%
 \endgroup
 \patchcmd{\titleblock@produce}
  {\frontmatter@RRAPformat}
  {\frontmatter@RRAPformat{\produce@RRAP{*#1\href{mailto:#2}{#2}}}\frontmatter@RRAPformat}
  {}{}
}%
\makeatother
\begin{document}

\preprint{AIP/123-QED}

\title[Waveguided sources of consistent, single-temporal-mode squeezed light: the good, the bad, and the ugly.]{Waveguided sources of consistent, single-temporal-mode squeezed light:\\ the good, the bad, and the ugly.}
\author{Martin Houde}
 \email{martin.houde@polymtl.ca.}
\author{Nicolás~Quesada}%
 \email{nicolas.quesada@polymtl.ca.}
\affiliation{ 
Department of Engineering Physics, École Polytechnique de Montréal, 2500 Chem. de Polytechnique, Montréal, Quebec H3T 1J4, Canada}%

\date{\today}

\begin{abstract}
	We study theoretically how the brightness of the pumps, with fixed profiles, affects the temporal mode structure of squeezed states generated by fixed parametric waveguided sources. We find that the temporal modes of these squeezed states can be partially mismatched and thus distinguishable, which is undesirable when using these states as resources for quantum computing or heralded state generation. By studying common frequency filtering techniques used experimentally, we find that although one can regain indistinguishability it comes at the price of potentially greatly reducing the purity of the state. We consider three different source configurations: unapodized single pass, apodized single pass, and apodized double pass. We find that the double pass configuration produces optimal results with almost perfectly indistinguishable states over varying degrees of brightness.
\end{abstract}

\maketitle
	\section{Introduction}
	\label{sec:Intro}

	Squeezed states of light are an important resource for quantum information and quantum computing ~\cite{braunstein2005info,Weedbrook2012info,braunstein2005squeezing}. In recent years, there has been considerable theoretical~\cite{christ2013theory,christ2014theory,Helt2020Degen,quesada2022BPP,quesada2020theory,vernon2019scalable, arzani2018versatile} and experimental ~\cite{zhong2020exp1,zhong2021exp2,zhong2019experimental,eckstein2011highly,wang2019gbs,Deshpande2022advantage,Arrazola2021circuits,triginer2020cascaded,harder2013optimized,harder2016single} advancements made into how such states are generated and used as a resource to show quantum computational advantage in the optical regime. Indeed, Gaussian boson sampling (GBS)~~\cite{hamilton2017gbs,kruse2019detailed,quesada2018gaussian,Deshpande2022advantage,grier2021complexity} has emerged as a candidate to show quantum computational advantage using squeezed states.  
	
	In GBS, one sends a set of spectrally identical input squeezed states into an interferometer and then measures the output photon distribution ~\cite{hamilton2017gbs,quesada2018gaussian,grier2021complexity}. Sampling from the theoretical probability distribution of the resulting detection patterns has been shown to be a computationally hard task ~\cite{hamilton2017gbs,Deshpande2022advantage,Bulmer2022advantage} and as such an ideal GBS experiment would show a quantum computational advantage. Such an advantage relies heavily on the indistinguishability of the input squeezed states. If the input states are partially distinguishable, the theoretical probability distribution simplifies and it becomes easier to sample from it classically therefore nullifying any quantum computational advantage~\cite{Valery2022distinguish,Shi2021distinguish}.
	
	As outlined in Ref.~\onlinecite{vernon2019scalable}, for a source of squeezed light to be considered practically useful for continuous variable quantum sampling and heralding it must meet several criteria: easily scalable, produce light in a single temporal mode across multiple sources and a wide range of squeezing levels, produce sufficiently high levels of squeezing, and be compatible with single and photon-number resolving detectors. In this paper, we focus mainly on the second criterion and set out to study how variations in pump brightness can lead to partially distinguishable states when considering identical nonlinear media and pump profiles.
	
\begin{figure}[t]
		\includegraphics[width=\linewidth]{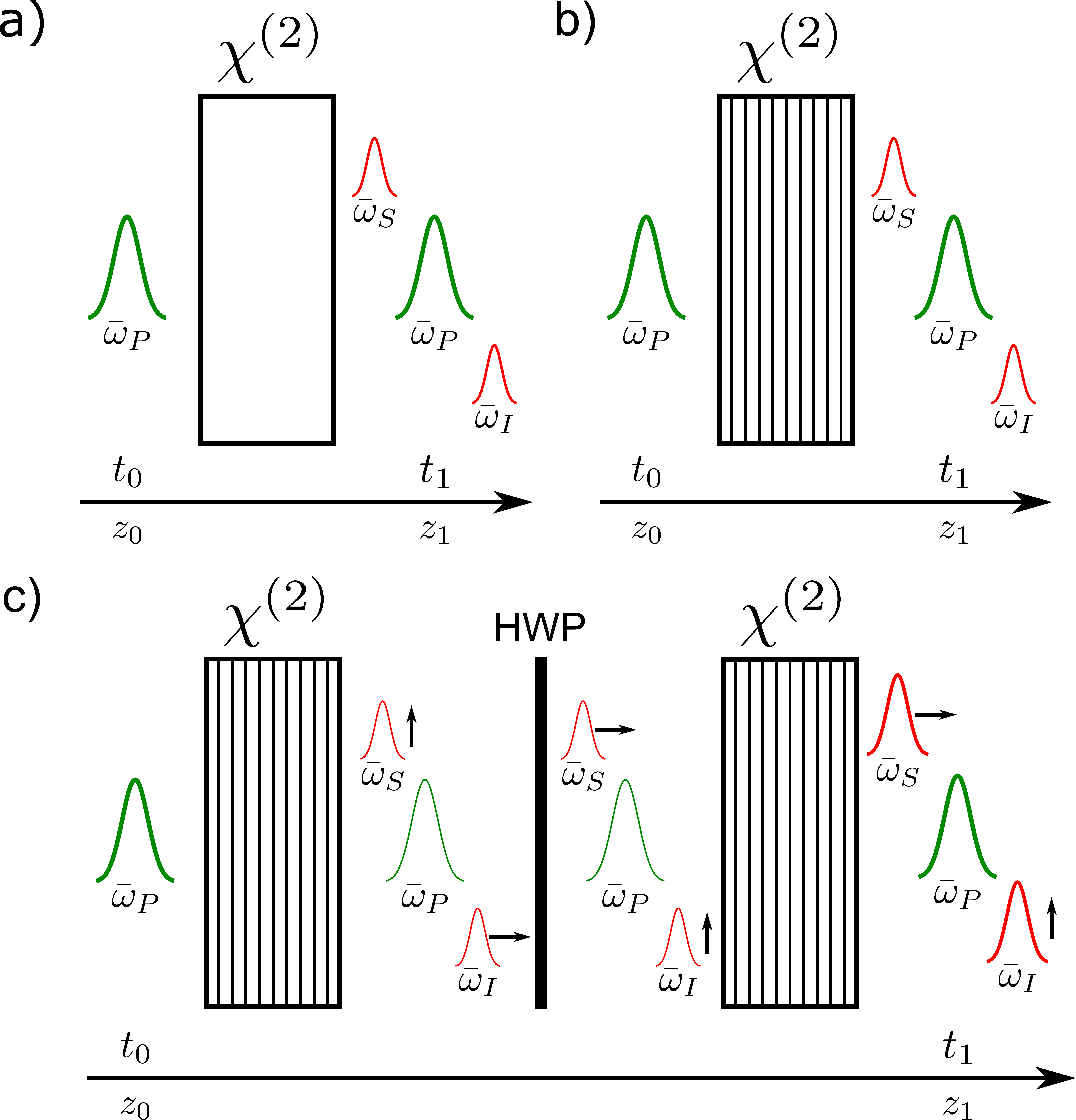}
		\caption{Propagation setup for type-\RomanNumeralCaps{2} SPDC. a) Unapodized single pass structure where a pump field localized at $z_{0}$ and central frequency $\bar{\omega}_{P}$ is sent towards an unapodized nonlinear region. Upon leaving the region, twin-beams are created at central frequencies $\bar{\omega}_{S}$ and $\bar{\omega}_{I}$.b) Apodized single pass structure where the nonlinear region is domain engineered(represented by black lines) to give a Gaussian phase-matching function (PMF).c) Apodized double pass structure where after the first pass, the signal and idler modes are polarization swapped(indicated by the black arrows) via a half-wave plate (HWP) before passing through a second nonlinear region. Both nonlinear regions are apodized to give a Gaussian PMF. In all configurations we assume no reflections at any of the interfaces.}
		\label{fig:model}
	\end{figure}

	We study three different setups of squeezed state generation. In the first configuration setup, shown in Fig.~\ref{fig:model}(a), the pump mode is sent through a single flat or unapodized nonlinear region and the squeezed state is analyzed at the end of said region. In the second configuration, shown in Fig.~\ref{fig:model}(b), we consider the case where the nonlinearity profile is domain engineered by aperiodic poling to apodize the phase-matching function~~\cite{branczyk2010optimized,dixon2013spectral,tambasco2016domain,dosseva2016shaping}. In the third configuration corresponding to the setup in Fig.~\ref{fig:model}(c), we take the output of the first setup with a domain engineered nonlinear region and pass it through a half-wave plate (HWP) before sending it through a second nonlinear region of the same length and strength but a flipped domain engineering. One could equally use a mirror to reflect the modes back through the same nonlinear region, as is done in Ref.~~\onlinecite{zhong2021exp2,lamas2001stimulated,Eisenberg2004doublepass}. The HWP swaps the polarizations of the signal and idler modes. We then analyze the squeezed state at the end of the second region. In both setups, we assume that reflections at the interfaces of the nonlinear regions are negligible, as is the case in the experiments of Ref.~~\onlinecite{zhong2020exp1, zhong2021exp2}. If reflections occur, one needs to consider a more sophisticated formalism as presented in Ref.~~\onlinecite{Liscindi2012Asymptotic}.
	
	Typical parametric waveguided sources can allow for both spontaneous parametric down conversion (SPDC) and four-wave mixing processes~\cite{quesada2022BPP}, both of which can be used to generate squeezed states. In this paper, we focus on SPDC processes which generate signal and idler modes of orthogonal polarizations (known as type-\RomanNumeralCaps{2} SPDC). For the scope of this paper, where we are interested in the effects of varying pump brightness on the generated squeezed state, we ignore self and cross phase modulation effects which are very small in the limit that we have a large classical pump and at most tens of photons are created~~\cite{triginer2020cascaded}. As such, we will only be focusing on $\chi^{2}$ interactions. We also ignore loss, which could be different for both signal and idler modes, however, as this is a Gaussian process it would be easy to implement in our model. Furthermore, we focus solely on the temporal mode structure along the direction of propagation and ignore the degrees of freedom of the transverse modes. If the longitudinal modes are indistinguishable, GBS protocols exists for both distinguishable~~\cite{grier2021complexity} and indistinguishable~~\cite{hamilton2017gbs,quesada2018gaussian} transverse modes and so these modes need not be considered. Furthermore, other useful protocols such as heralding also only focus on the longitudinal modes~\cite{tiedau2019scalability, engelkemeier2021climbing,branczyk2010optimized,blay2017effects,meyer2017limits,thomas2021general}.    
	
	By considering the first setup with an unapodized nonlinear region and analyzing the joint spectral amplitude(JSA) and the Schmidt number for different pump intensities, we show that the output state is always spectrally mixed. We study the temporal mode structure and show that different pump intensities lead to noticeably mismatched modes. In fact, we show that the fidelity between two modes generated by different pump intensities can drop to below $90\%$ when considering a low-gain level mode and a mode with approximately ten photons. Although filtering can be used to increase spectral purity, the output state is never properly described by a single temporal mode. Furthermore, the filtered temporal modes, which are thermal squeezed states, are described by distinguishable squeezed and thermal parts. An unapodized nonlinear region therefore does not satisfy the stated criteria at all.  
	
	In the first setup using an apodized nonlinear region, by analyzing the JSA for different pump intensities, we show that high-gain effects lead to an apparent increase in frequency correlations. However, by studying the Schmidt number, we find that the output state remains relatively spectrally pure for different intensities. We study the temporal mode structure and again show that different pump intensities lead to noticeably mismatched modes. As in the unapodized case, we show that the fidelity between two modes generated by different pump intensities can drop to below $90\%$ when considering a low-gain level mode and a mode with roughly ten photons.  Furthermore, we show that even though we can decrease spectral distinguishability by filtering, it comes at the cost of reducing the state purity. This state purity loss is much more pronounced for states generated by stronger pumps. Although much better than the unapodized case, a single apodized nonlinear region still does not fully satisfy the stated criteria.
	
	Finally, we do the same analysis for the double pass setup and show that it performs much better. For the same levels of gain as the first setup, we show that the high-gain effects are not as detrimental to the JSA and spectral purity. We also show that the temporal mode mismatch is not as pronounced. By studying the fidelity between modes generated by different pump intensities, we show that it only drops by slightly less than $1\%$. Since the states generated in this setup are much less distinguishable, we do not study the effects of filtering. From our analysis, we find that the double pass structure fulfills the criteria required for practically useful continuous variable quantum computing. 
	
	The paper is organized as follows. In Sec.~\ref{sec:Model} we derive the equations of motion for the signal and idler modes for type-\RomanNumeralCaps{2} SPDC and detail how to obtain numerical solutions. In Sec.~\ref{sec:Filt} we develop a model to study filtered output states when the input squeezed states are either spectrally pure or mixed. In Sec.~\ref{sec:Unpoled}, we study the results for the setup of Fig.~\ref{fig:model}(a) with an unapodized nonlinear region. Since the squeezed output states are spectrally mixed, we study how filtering modifies the outputs. In Sec.~\ref{sec:SPR}, we study results for the setup of Fig.~\ref{fig:model}(b) with an apodized nonlinear region. Unlike the unapodized case, the squeezed output states are approximately spectrally pure but distinguishable. As such, we also study how filtering affects the outputs. In Sec.~\ref{sec:DP}, we study results for the setup of Fig.~\ref{fig:model}(c) with apodized nonlinear regions. Since the double pass setup gives much better results than the single pass, we do not study filtering techniques. Finally, in Sec.~\ref{sec:conc} we conclude our findings.
	
	\section{Model of Twin-Beam generation}
	\label{sec:Model}
	
	We follow the twin-beam dynamics of Ref.~~\onlinecite{quesada2020theory} and leave a detailed derivation in Appendix~\ref{app:derivation}.  
	
	Without loss of generality, we assume that the propagation of the modes is in the $z$ direction. For each mode, $j=P,S,I$ for pump, signal, and idler respectively, we associate a central wavevector $\bar{k}_{j}$ with a central frequency $\bar{\omega}_{j} = \omega_{j,\bar{k}_{j}}$ and assume a linear dispersion
	\begin{align}\label{Eq:DispersionMT}
		\omega_{j,k} \approx \bar{\omega}_{j} +v_{j}(k-\bar{k}_{j}),
	\end{align}
	where the group velocity $v_{j}$ is taken to be constant over the frequency ranges of interest and we ignore group velocity dispersion within each frequency range. As we are interested in type-\RomanNumeralCaps{2} SPDC processes, we require that
	\begin{align}
		\bar{\omega}_{P}-\bar{\omega}_{S}-\bar{\omega}_{I}&=0,\\
		\bar{k}_{P}-\bar{k}_{S}-\bar{k}_{I}&=0.\label{Eq:MomconMT}
	\end{align}
	Note that if quasi-phase matching is used, the right-hand side of Eq.~\ref{Eq:MomconMT} should be changed to $\pm 2\pi/\Lambda_{\text{pol}}$ where $\Lambda_{\text{pol}}$ is the poling period. Furthermore, we assume self and cross phase modulation terms to be negligible and assume that the pump mode is prepared in a strong coherent state with a large number of photons which remains constant throughout the interaction (undepleted-classical pump approximation). Under these assumptions, we obtain equations for the spatial evolution of the signal and idler operators
	\begin{align}
		\frac{\partial}{\partial z}a_{S}(z,\omega) =& i\Delta k_{S}(\omega)a_{S}(z,\omega)\label{Eq:SigEomZ}\\
		&+i\frac{\gamma g(z)}{\sqrt{2\pi}}\int d\omega' \beta_{P}(z,\omega+\omega')a^{\dagger}_{I}(z,\omega'),\nonumber\\
		\frac{\partial}{\partial z}a^{\dagger}_{I}(z,\omega) =& -i\Delta k_{I}(\omega)a^{\dagger}_{I}(z,\omega)\label{Eq:IdEomZ}\\
		&-i\frac{\gamma^{*} g(z)}{\sqrt{2\pi}}\int d\omega' \beta^{*}_{P}(z,\omega+\omega')a_{S}(z,\omega'),\nonumber
	\end{align}
	where in the first terms on the right-hand side we have
	\begin{align}
		\Delta k_{S}(\omega)=\left(\frac{1}{v_{S}} - \frac{1}{v_{P}}\right)(\omega_{S}-\bar{\omega}_{I}),\\
        \Delta k_{I}(\omega)=\left(\frac{1}{v_{I}} - \frac{1}{v_{P}}\right)(\omega_{I}-\bar{\omega}_{I}).
	\end{align}
	For the second term on the right-hand side, responsible for twin-beam generation, we introduce the coupling parameter given by
	\begin{align}
		\gamma g(z) = \frac{\xi (z)}{\sqrt{\hbar\bar{\omega}_{P}v_{P}v_{S}v_{I}}},
	\end{align}
	where $g(z)$ is the poling function with $g(z)=0$ where the nonlinearity is absent and either 1 or -1 depending on the orientation of the nonlinear region(see Appendix~\ref{app:derivation} for details on $\xi(z)$ which is the nonlinear coupling parameter for second-order interactions). The spectral content of the pump, $\beta_{P}(z,\omega)=\beta_{P}(\omega)$, which appears in the integrals is independent of $z$ under our assumptions.
	
	As shown in Ref.~~\onlinecite{quesada2020theory}, these operators obey canonical bosonic commutation relations for all $z$ and we can interpret quantities such as $a^{\dagger}_{l}(z,\omega)a_{l}(z,\omega)$ as a photon frequency density operator at position $z$.


	\subsection{Solving the equations of motion}
	
	To solve numerically, we begin by discretizing the operators $a_{j}(z,\omega)$ in frequency space such that $\omega_{n} = \omega_{0}+n\Delta\omega|^{N-1}_{0}$ for an $N$-size grid. To simplify notation, we introduce column vectors
	\begin{align}
		\bm{u}_{n}(z) = a_{S}(z,\omega_{n}),\\
		\bm{v}^{\dagger}_{n}(z) = a^{\dagger}_{I}(z,\omega_{n}),
	\end{align}
	which allows us to re-write the equations of motion (Eq.~\ref{Eq:SigEomZ}, \ref{Eq:IdEomZ}) in block-matrix form
	\begin{align}
		\frac{\partial}{\partial z}\begin{pmatrix}
			\bm{u}(z)\\
			\bm{v}^{\dagger}(z)
		\end{pmatrix} &= i\begin{bmatrix}
			\bm{G}(z) & \bm{F}(z) \\
			-\bm{F}^{\dagger}(z) & -\bm{H}^{\dagger}(z)
		\end{bmatrix}\begin{pmatrix}
			\bm{u}(z)\\
			\bm{v}^{\dagger}(z)
		\end{pmatrix}\\
		&= i\bm{Q}(z)\begin{pmatrix}
			\bm{u}(z)\\
			\bm{v}^{\dagger}(z)
		\end{pmatrix}\nonumber,
	\end{align}
	where the matrix elements are given by
	\begin{align}
		\bm{G}_{n,m}(z) &= \Delta k_{S}(\omega_{n})\delta_{n,m},\\
		\bm{H}_{n,m}(z) &= \Delta k_{I}(\omega_{n})\delta_{n,m},\\
		\bm{F}_{n,m}(z) &= \frac{\gamma g(z)}{\sqrt{2\pi}}\beta_{P}(\omega_{n}+\omega_{m})\Delta\omega.
	\end{align}
	The only $z$-dependence of the $\bm{Q}(z)$ matrix comes from $g(z)$ which depends on the given poling of the nonlinear medium. Experimentally and theoretically, the poling function $g(z)$ will take on values $+1$ or $-1$ for short sections of length $\Delta z$. For each short section, the equations of motion are $z$-independent and can be readily solved via matrix exponentiation. Indeed, for a given section with known value of $g(z)$, the solution is given by
	\begin{align}
		\begin{pmatrix}
			\bm{u}(z_{0}+\Delta z)\\
			\bm{v}^{\dagger}(z_{0}+\Delta z)
		\end{pmatrix}= \bm{U}(z_{0}+\Delta z,z_{0})\begin{pmatrix}
			\bm{u}(z_{0})\\
			\bm{v}^{\dagger}(z_{0})\end{pmatrix}.\label{Eq:Soln}
	\end{align}
	The propagator $\bm{U}(z_{0}+\Delta z,z_{0})$ is given by
	\begin{align}
		&\bm{U}(z_{0}+\Delta z,z_{0}) = \text{exp}\left(i\Delta z \bm{Q}(z_{0})\right)\\
		&= \begin{bmatrix}
			\bm{U}^{S,S}(z_{0}+\Delta z,z_{0}) &  \bm{U}^{S,I}(z_{0}+\Delta z,z_{0})\\
			\left(\bm{U}^{I,S}(z_{0}+\Delta z,z_{0})\right)^{*} & \left(\bm{U}^{I,I}(z_{0}+\Delta z,z_{0})\right)^{*}
		\end{bmatrix}
	\end{align}
	where it is understood that $g(z_{0})$ in $\bm{Q}(z_{0})$ takes the given value over the short section spanning $\left( z_{0},z_{0} +\Delta z \right)$.
	
	For a known poling function, $g(z)$, we can form a complete solution by stitching together the solutions over every section $\Delta z$. The formal solution is given by
	\begin{align}
		\bm{U}(z,z_{0})=\prod_{p}\text{exp}\left(i\Delta z_{p}\bm{Q}(z_{p})    \right)
	\end{align}
	where different segments can possibly be of different lengths. Although this might look numerically heavy to compute, we can optimize it in practice. For simplicity, we choose the segments to be of equal length. Choosing a $\Delta z$ small enough will always guarantee that we can approximate any experimental poling very well. When all segments are of equal lengths we only need to calculate two matrix exponentiations(for each sign of $g(z)$). We find that we can then optimize the stitching of the full solution by first precalculating all possible configurations of four domains and then stitching the full solution in chunks of four domains at a time rather than going one domain at a time.
	
	Taking the continuous form of Eq.~\ref{Eq:Soln}, we find the solution for the signal and idler operators to be
	\begin{align}
		a_{S}(z,\omega) =& \int d\omega' U^{S,S}(\omega,\omega';z,z_{0})a_{S}(z_{0},\omega')\label{eq:SigSolnCont}\\
		&+\int d\omega' U^{S,I}(\omega,\omega';z,z_{0})a^{\dagger}_{I}(z_{0},\omega'),\nonumber\\
		a^{\dagger}_{I}(z,\omega') = & \int d\omega' \left(U^{I,S}(\omega,\omega';z,z_{0})\right)^{*}a_{S}(z_{0},\omega')\label{eq:IdSolnCont}\\
		&+ \int d\omega' \left(U^{I,I}(\omega,\omega';z,z_{0})\right)^{*}a^{\dagger}_{I}(z_{0},\omega'),\nonumber
	\end{align}
	where the continuous propagator blocks are related to the discretized blocks such that
	\begin{align}
		U^{i,j}(\omega_{n},\omega_{m};z,z_{0}) =\bm{U}^{i,j}_{n,m}(z,z_{0})/\Delta\omega.
	\end{align}
	
	We can now relate the operators at the end of the nonlinear region to those at the beginning. As we will be interested in the case where $z=z_{1}$ (see Fig.~\ref{fig:model}), we introduce input and output operators $a^{(\text{in/out})}_{l}(\omega)=e^{-i\Delta k_{l}(\omega) z_{0/1}}a_{l}(z_{0/1},\omega)$, for both signal and idler modes, which allows us to omit the spatial dependence from here on and also removes the trivial spatial evolution (i.e. phase-factors from free propagation).

	Given the linearity of the solution (Eqs.~\ref{eq:SigSolnCont},\ref{eq:IdSolnCont}), the state we obtain by applying the operators on the vacuum is Gaussian and is fully described by its first and second moments. The first moments are all zero and the only non-vanishing second moments are
	\begin{align}
		N_{l}(\omega,\omega') &= \langle \text{vac}| a^{(\text{out})\dagger}_{l}(\omega)a^{(\text{out})}_{l}(\omega')   |\text{vac}\rangle\nonumber\\
		&= \sum_{\lambda}\sinh^{2}(r_{\lambda})\left(\rho^{l}_{\lambda}(\omega)  \right)^{*}\rho^{l}_{\lambda}(\omega'),\label{Eq:Numcor}\\
		M(\omega,\omega')&=\langle \text{vac}| a^{(\text{out})}_{S}(\omega)a^{(\text{out})}_{I}(\omega')   |\text{vac}\rangle\nonumber\\
		&=\sum_{\lambda}\frac{\sinh(2r_{\lambda})}{2}\rho^{S}_{\lambda}(\omega)\rho^{I}_{\lambda}(\omega'),\label{Eq:Mcor}
	\end{align}
	where again $l=S,I$ and the $r_{\lambda}$'s are the squeezing parameters associated with the Schmidt modes \{$\rho^{S/I}_{\lambda}$\}. The Schmidt modes, which give the temporal structure of the squeezed state, are obtained from the singular value decompositions of the transfer functions in Eqs.~\ref{eq:SigSolnCont},~\ref{eq:IdSolnCont}~\cite{quesada2020theory}. We can also characterize the level of gain by considering the average number of signal photons
	\begin{align}
		\langle N_{S}\rangle = \int d\omega N_{S}(\omega, \omega)=\sum_{\lambda} \sinh^2 (r_\lambda). \label{Eq:Ns}
	\end{align}
	
	From the squeezing parameters, we can study the spectral purity of the squeezed output state by considering how the Schmidt number
	\begin{align}\label{Eq:schmidt}
		K = \frac{\left( \sum_{\lambda}\sinh^{2}(r_{\lambda})\right)^{2}}{\sum_{\lambda}\sinh^{4}(r_{\lambda})}
	\end{align}
	evolves as a function of gain~\cite{Christ2011probing}. Recall that the spectral purity is a measure of  frequency correlations in the JSA and is not related to the state purity of the density matrix. For a spectrally pure state, we have $K=1$ whereas $K>1$ otherwise.
	
	Given these moments, as outlined in ~\cite{quesada2022BPP,quesada2020theory}, one can also easily construct the JSA
	\begin{align}
		J(\omega,\omega') = \sum_{\lambda}r_{\lambda}\rho^{S}_{\lambda}(\omega)\rho^{I}_{\lambda}(\omega').
	\end{align}
	It is well known that in the low-gain limit, the JSA can be approximated as
	\begin{align}\label{eq:JSA}
		J(\omega,\omega')\propto \Phi\left[\Delta k_{S}(\omega)+\Delta k_{I}(\omega')\right]\beta_{P}(\omega+\omega')
	\end{align}
	where $ \Phi\left[\Delta k_{S}(\omega)+\Delta k_{I}(\omega')\right]$ is the phase-matching function and $\beta_{P}(\omega+\omega')$ is the spectral content of the pump. Furthermore, the phase-matching function is related to the poling function $g(z)$:
	\begin{align}
		\Phi\left[\Delta k_{S}(\omega)+\Delta k_{I}(\omega')\right] \propto \int \frac{dz}{\sqrt{2\pi}}e^{-iz\left(  \Delta k_{S}(\omega)+\Delta k_{I}(\omega') \right)}g(z).
	\end{align}
	For an unapodized region, the phase-matching function is a Sinc function ~\cite{quesada2020theory}. As outlined in ~\cite{branczyk2010optimized,dixon2013spectral,dosseva2016shaping,tambasco2016domain,Aggie2021domain} and shown experimentally in ~\cite{zhong2020exp1,zhong2021exp2,harder2013optimized, harder2016single,triginer2020cascaded,xin2022spectrally}, it is possible to find a poling function that gives rise to a Gaussian phase-matching function. This in turn leads to a separable JSA in the low-gain regime~\cite{u2006generation,quesadaAggie2018optimal}; a property that is well sought out~\cite{quesada2022BPP,quesadaAggie2018optimal}.
	Note that besides poling, one can also take advantage of the tensor properties of the nonlinear response to engineer a Gaussian PMF~~\cite{poveda2022custom}.
	The squeezed state at the output of the nonlinear region is given by the Hilbert space ket
	\begin{align}
		&|\Psi \rangle =\nonumber\\ &\text{exp}\left(\int d\omega d\omega' J(\omega,\omega')a^{(\text{in})\dagger}_{S}(\omega)a^{(\text{in})\dagger}_{I}(\omega')-\text{H.c.}    \right)|\text{vac}\rangle.
	\end{align}
	Note that the state above has purity one (since it is a pure state) but can have any spectral purity, which will depend on the frequency correlation of the JSA $J(\omega,\omega')$.
	
	\section{Filtering}
	\label{sec:Filt}
	Filtering is a common method used to increase spectral purity and the usefulness of low-gain squeezed states~\cite{christ2014theory,branczyk2010optimized,thomas2021general,blay2017effects,meyer2017limits}. Filtering adds mixedness to the state and can be thought of as adding frequency dependent loss via beamsplitter transformations for each mode such that the operator of the filtered mode transforms as
	\begin{align}\label{Eq:beamsplitter}
		\tilde{a}_{l}(\omega) = T_{l}(\omega)a^{(\text{out})}_{l}(\omega)+\sqrt{1-|T_{l}(\omega)|^{2}}\zeta_{l}(\omega).
	\end{align}
	The index $l=S,I$ for signal and idler mode respectively, $T_{l}(\omega)$ is the transmission coefficient, and $\zeta_{l}$ is a bath operator. As this transformation is linear, the filtered state is also Gaussian.  We assume symmetric filtering where $T_{S}(\omega)=T_{I}(\omega)=T(\omega)$. As such, we can describe the filtered output states as a general thermal squeezed state (th-SS) with density matrix
	\begin{align}\label{Eq:thermal}
		\varrho_{\text{th-SS}} = S \left\{ \bigotimes_{i}\left[\varrho^{\text{th}}_{\mathcal{A}_{i}} \otimes \varrho^{\text{th}}_{\mathcal{B}_{i}}\right]\right\}S^{\dagger}
	\end{align}
	where
	$
	S=\text{exp}\left[\sum_{\lambda}r_{\lambda}A^{\dagger}_{\lambda}B^{\dagger}_{\lambda}- \text{H.c.}   \right]
	$
	is the squeezing operator for the filtered Schmidt modes $A_{\lambda}, B_{\lambda}$ with squeezing parameter $r_{\lambda}$, and $\varrho^{\text{th}}_{\mathcal{C}_{i}}(\bar{n}_{i})=\exp\left[-\beta_{i}\mathcal{C}^{\dagger}_{i}\mathcal{C}_{i}  \right]/\text{Tr}\left(\exp\left[-\beta_{i}\mathcal{C}^{\dagger}_{i}\mathcal{C}_{i}  \right] \right)$ describes a single-mode thermal state, $\mathcal{C}_{i} \in \{ \mathcal{A}_{i}, \mathcal{B}_{i} \}$ are broadband operators satisfying the canonical commutation relations $[\mathcal{A}_{i}, \mathcal{A}_{j}^\dagger] = [\mathcal{B}_{i}, \mathcal{B}_{j}^\dagger] = \delta_{ij}  $
	implying that their associated temporal mode amplitudes are orthogonal. The average photon number is given by $\bar{n}_{i}=\left( e^{\beta_i} - 1 \right)^{-1}$. When working in the symmetric group-velocity-matched regime where the JSA is symmetric and since the filtering is the same for both signal and idler modes, the $\mathcal{A}_{i}$ and $\mathcal{B}_{i}$ modes have the same average photon numbers and profiles. The modes describing the thermal distribution are in general not the same as the modes describing the squeezing. In fact, the two sets of modes coincide only when the pre-filtered squeezed state is spectrally pure (e.g. $K=1$) as we will show below.

	\subsection{Spectral Purity and Filtering}
	Since the filtered output state is Gaussian, we can fully describe it using its covariance  matrix which consequently depends solely on the second moments of the filtered operators $\tilde{a}_{i}(\omega)$. Using Eq.~\ref{Eq:beamsplitter} and the unfiltered correlators of Eqs.~\ref{Eq:Numcor} and \ref{Eq:Mcor} we find that
	\begin{align}
		\tilde{N}_{l}(\omega,\omega') &= T^{*}(\omega)N_{l}(\omega,\omega')T(\omega'),\\
		\tilde{M}(\omega,\omega') &= T(\omega)M(\omega,\omega')T(\omega').
	\end{align}
	In terms of the Schmidt modes introduced earlier, these correlators can be expressed as
	\begin{align}
		\tilde{N}_{l}(\omega,\omega') &= \sum_{\lambda} \sinh^{2}{(r_{\lambda})} \left(T(\omega)\rho^{l}_{\lambda}(\omega)\right)^{*}T(\omega')\rho^{l}_{\lambda}(\omega'),\label{Eq:NumcorF}\\
		\tilde{M}(\omega,\omega') &=\sum_{\lambda}\frac{\sinh{(2r_{\lambda})}}{2}T(\omega)\rho^{S}_{\lambda}(\omega)T(\omega')\rho^{I}_{\lambda}(\omega').
	\end{align}
	We now consider the two different cases where $K=1$ and $K>1$ to see what the proper modes are to describe the filtered state.
	
	\subsubsection{Spectrally pure}\label{sec:pure}
	When $K=1$, the squeezed output state is spectrally pure and is thus described by a single Schmidt mode. In this case, one can easily identify the new temporal modes that describe the filtered state. From Eq.~\ref{Eq:NumcorF}, with $\lambda = 1$, we can identify new, unnormalized, Schmidt modes $\tilde{A}^{l}_{1}(\omega) = T(\omega)\rho^{l}_{1}(\omega)$. Letting the normalization constant be $\eta^{l}_{1}=\int d\omega T(\omega)\rho^{l}_{1}(\omega) \left( T(\omega)\rho^{l}_{1}(\omega) \right)^{*}$ and defining normalized modes $A^{l}_{1}(\omega) = \tilde{A}^{l}_{1}(\omega)/\sqrt{\eta_{l}}$ the correlators take the form
	\begin{align}
		\tilde{N}_{l}(\omega,\omega') &= \eta^{l}_{1}\sinh^{2}(r_{1})\left(A^{l}_{1}(\omega)\right)^{*}A^{l}_{1}(\omega')\label{eq:pureN},\\
		\tilde{M}(\omega,\omega') &=\sqrt{\eta^{S}_{1}\eta^{I}_{1}}\frac{\sinh{(2r_{1})}}{2}A^{S}_{1}(\omega)A^{I}_{1}(\omega)\label{eq:pureM}.
	\end{align}
	The temporal mode structure of the filtered state is given by the Schmidt modes $A^{S/I}_{1}(\omega)$ that have undergone transmission loss. Since the correlators depend solely on these modes, the modes describing the thermal distribution in Eq.~\ref{Eq:thermal} are exactly these Schmidt modes.
	
	In this case, we can obtain the filtered state's temporal structure by simply applying the filter function to the Schmidt mode of the squeezed output state and renormalizing properly.
	
	\subsubsection{Spectrally mixed filtering}\label{sec:mixedfilt}
	When $K>1$, there is more than one relevant Schmidt mode. Naively, one might just generalize the $\lambda=1$ moments (see Eqs.~\ref{eq:pureN},\ref{eq:pureM}) such that
	\begin{align}
		\tilde{N}_{l}(\omega,\omega') &= \sum_{\lambda} \eta^{l}_{\lambda}\sinh^{2}{(r_{\lambda})} \left(A^{l}_{\lambda}(\omega)\right)^{*}A^{l}_{\lambda}(\omega'),\\
		\tilde{M}(\omega,\omega') &=\sum_{\lambda}\sqrt{\eta^{S}_{\lambda}\eta^{I}_{\lambda}}\frac{\sinh{(2r_{\lambda})}}{2}A^{S}_{\lambda}(\omega)A^{I}_{\lambda}(\omega),
	\end{align}
	however, this is not the proper solution. These new sets of Schmidt modes are no longer orthonormal and thus do not describe the temporal mode structure of the filtered state. This tells us that the thermal distribution modes in Eq.~\ref{Eq:thermal} are different modes.
	
	Finding the temporal mode structure in this case is not as simple as applying the filter function to the Schmidt modes of the squeezed output state. One additionally needs to find the set of thermal modes and their average occupancies.
	
	To do so, we consider the discretized system on a grid of size $N$, and the filtered output modes $\tilde{a}_{l}(\omega_{n})$ with $n\in [0,N-1]$. In terms of the hermitian phase quadratures $x_{l}(\omega_{n})$ and $p_{l}(\omega_{n})$ we can express the filtered output modes as $\tilde{a}_{l}(\omega_{n}) = \left(x_{l}(\omega_{n})+ip_{l}(\omega_{n})   \right)/\sqrt{2\hbar}$. We define a row vector $\bm{r}^{T} = \left(\bm{x}_{S},\bm{x}_{I},\bm{p}_{S},\bm{p}_{I}\right)$ where $\bm{x}_{S/I,n}~=~x_{S/I}(\omega_{n})$ and $\bm{p}_{S/I,n}=p_{S/I}(\omega_{n})$. Doing so allows us to express the covariance matrix of the filtered output state as $\bm{V} =\langle \{\bm{r},\bm{r}^{T} \}  \rangle/2 $ where $\{A,B  \}=AB+BA$ is the anti-commutator.

	\begin{figure*}[ht]
		\includegraphics[width=\linewidth]{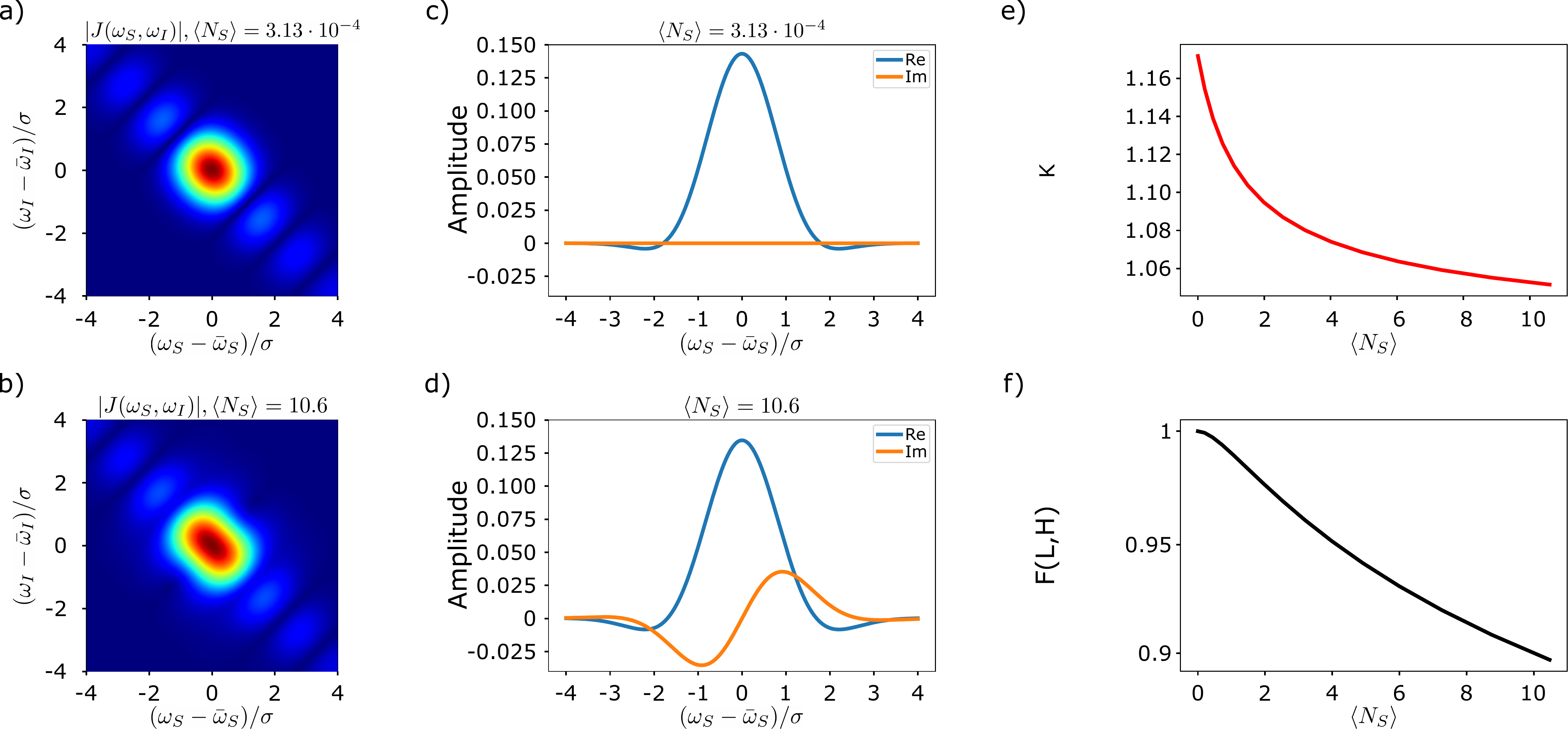}
		\caption{Results for the single pass setup with an unapodized nonlinear region (c.f. Sec.~\ref{sec:Unpoled}). a) Absolute value of the JSA is the low-gain regime with $\langle N_{S} \rangle = 3.13\cdot 10^{-4}$. As expected by the lack of poling, the JSA is approximately a product of a Gaussian and a Sinc. b) Absolute value of the JSA in the high-gain regime with $\langle N_{S} \rangle = 10.6$. High-gain effects leads to a warping of the JSA. c) First Schmidt mode, $\rho^{S}_{1}(\omega)$, in the low-gain regime with $\langle N_{S} \rangle = 3.13\cdot 10^{-4}$. d)  First Schmidt mode, $\rho^{S}_{1}(\omega)$, in the high-gain regime with $\langle N_{S} \rangle = 10.6$. In the high-gain regime, the first Schmidt mode has a much greater imaginary part telling us that the Schmidt modes at different gains are distinguishable. e) The Schmidt Number, $K$, as a function of gain, characterized by $\langle N_{S} \rangle$. The Schmidt number is always greater than unity and we are therefore always spectrally mixed. f) The fidelity, $F(L,H)$, between the first Schmidt mode at fixed low gain, $\rho^{S}_{L}(\omega)$, and the first Schmidt mode at variable gain, $\rho^{S}_{H}(\omega)$, as a function of gain. The fidelity drops to slightly below $90\%$ which indicates partial distinguishability of Schmidt modes at different levels of gain.}
		\label{fig:UnpoledJSA}
	\end{figure*}

	Since the filtered output state is fully described by its covariance matrix, we can obtain the relevant modes from it. The covariance matrix is a $4N\times 4N$ real definite matrix and as such can be decomposed as
	\begin{align}
		\bm{V} = \bm{S}\bm{D}\bm{S}^{T}
	\end{align}
	where $\bm{S}$ is a symplectic matrix and
	\begin{align}
		\bm{D} = \text{diag}(&\nu_{0},\nu_{0},\ldots,\nu_{N-1},\nu_{N-1},\nonumber\\
		&\nu_{0},\nu_{0},\ldots,\nu_{N-1},\nu_{N-1})    
	\end{align}
	(i.e. a diagonal matrix). We can relate the elements of the diagonal to thermal occupancies such that $\nu_{i} = \hbar(2\bar{n}_{i}+1)/2$. This is known as the Williamson decomposition ~\cite{serafiniBook}. Furthermore, we can decompose the symplectic matrix $\bm{S}$ as
	\begin{align}
		\bm{S} = \bm{O}\bm{\Lambda} \bm{\tilde{O}}^{T}
	\end{align}
	where $\bm{O}$ and $\bm{\tilde{O}}$ are both orthogonal and symplectic, and
	\begin{align}
		\bm{\Lambda}= \text{diag}(&e^{r_{0}},e^{r_{0}},\ldots,e^{r_{N-1}},e^{r_{N-1}},\nonumber\\
		&e^{-r_{0}},e^{-r_{0}},\ldots,e^{-r_{N-1}},e^{-r_{N-1}})
	\end{align}
	represents a set of pairs of single-mode squeezing operations with squeezing parameters $\{r_{0},\ldots, r_{N-1}  \}$. This is known as the Bloch-Messiah decomposition ~\cite{serafiniBook,Cariolar2016BM1,Cariolaro2016BM2,mccutcheon2018structure,horoshko2019bloch}. Since these are single-mode squeezing operations, we need to convert them into two-mode squeezing operations to properly describe the output state.
	
	We can combine two single-mode squeezed states, with opposite squeezing parameters, by passing them through a $50:50$ symmetric beamsplitter given by
	\begin{align}
		\bm{w} = \frac{1}{\sqrt{2}}\begin{pmatrix}
			1 & i \\
			i & 1
		\end{pmatrix}
	\end{align}
	which we can extend to $N$ pairs of single-mode squeezed state as
	\begin{align}
		\bm{B} = \bigoplus_{n=1}^{N}\bm{w}.
	\end{align}
	To transform the $\bm{\Lambda}$ matrix to two-mode squeezing operations, we need to act on it with
	\begin{align}
		\bm{W} = \begin{pmatrix}
			\text{Re}\left(\bm{B} \right) & \text{Im}\left(\bm{B} \right)\\
			-\text{Im}\left(\bm{B} \right) & \text{Re}\left(\bm{B} \right)
		\end{pmatrix}.
	\end{align}
	Putting everything together, we find that the covariance matrix can be decomposed to
	\begin{align}
		\bm{V} =& \bm{O}\bm{\Lambda}\bm{\tilde{O}}^{T}\bm{D}\bm{\tilde{O}}\bm{\Lambda}\bm{O}^{T}\nonumber\\
		=&\bm{O}\bm{W}^{T}\left[\bm{W}\bm{\Lambda}\bm{W}^{T}\right]\bm{W}\bm{\tilde{O}}^{T}\bm{D}\bm{\tilde{O}}\bm{W}^{T}\left[\bm{W}\bm{\Lambda}\bm{W}^{T}\right]\bm{W}\bm{O}^{T}\nonumber\\
		=& \bm{O}\bm{W}^{T}\bm{\Sigma}\bm{W}\bm{\tilde{O}}^{T}\bm{D}\bm{\tilde{O}}\bm{W}\bm{\Sigma}\bm{W}^{T}\bm{O}^{T}\nonumber\\
		=&\left( \bm{O}\bm{W}^{T} \right)\bm{\Sigma} \left(\bm{W}\bm{O}^{T} \right)\nonumber\\&\cdot\left(\bm{O}\bm{\tilde{O}}^{T}  \right)\bm{D}\left(\bm{\tilde{O}}\bm{O}^{T}  \right)\nonumber\\&\cdot\left( \bm{O}\bm{W}^{T} \right)\bm{\Sigma}\left(\bm{W}\bm{O}^{T}  \right)\label{Eq:Covmat}
	\end{align}
	where in the second line we have used $\mathbb{1} = \bm{W}^{T}\bm{W}$, in the third line we have introduced $\bm{\Sigma} = \bm{W}\bm{\Lambda}\bm{W}^{T}$ which now represents sets of two-mode squeezing operations, and in the last line we have used $\mathbb{1} = \bm{O}^{T}\bm{O}$ to obtain a form where the modes can be identified clearly.
	
	Note that the final expression of the covariance matrix in Eq.~\ref{Eq:Covmat} is of the form given in Eq.~\ref{Eq:thermal}. Therefore, the columns of symplectic-orthogonal matrix $\bm{O}\bm{W}^{T}$ tells us the real and imaginary parts of the filtered Schmidt modes corresponding \{$A_{\lambda}, B_{\lambda}$\} while the columns of the symplectic-orthogonal matrix $\bm{O}\tilde{\bm{O}}^{T}$ tells us the real and imaginary parts corresponding to the thermal modes \{$\mathcal{A}_{\lambda}, \mathcal{B}_{\lambda}$\}.
	We have implemented the numerical evolution of the squeezed states, as described in Sec.~\ref{sec:Model}, as well as their decompositions using the Symplectic formalism, as described in this section, in the Python library \texttt{NeedALight}\cite{needalight}.

	\section{Unapodized Single Pass Results}\label{sec:Unpoled}
	We begin by considering the worst case scenario for squeezed state generation: a nonlinear region corresponding to an unapodized PMF as in the configuration shown in Fig.~\ref{fig:model}(a). We consider a nonlinear region of length $l$ which allows for type-\RomanNumeralCaps{2} SPDC processes where photons from a pump mode are converted into photons in signal and idler modes of degenerate frequencies but orthogonal polarizations. We take the pump mode to be Gaussian with spectral content
	\begin{align}\label{Eq:pumpform}
		\beta_{P}(\omega) = \frac{\sqrt{\hbar \bar{\omega}_{P} N_{P}}}{\sqrt[4]{\pi\sigma^{2}}}\exp\left(-\frac{(\omega-\bar{\omega}_{P})^{2}}{2\sigma^{2}}\right),
	\end{align}
	which has mean bandwidth $\sigma$, and mean photon number $N_{P}$. We also work in the symmetric group-velocity-matched regime where
	\begin{align}\label{Eq:symgrpvel}
		\left( \frac{1}{v_{S}}-\frac{1}{v_{P}}  \right) = -\left( \frac{1}{v_{I}}-\frac{1}{v_{P}}  \right)
	\end{align}
	which gives rise to a symmetric JSA.  
	
	We follow the evolution of the generated light for arbitrary gain, which we characterize by the mean number of signal photons $\langle N_{S}\rangle$ given by Eq.~\ref{Eq:Ns},  by tuning the mean number of pump photons $N_{P}$. We study the squeezed output states for different levels of gain without modifying the nonlinear region nor the envelope of the pump. To solve numerically, we break the frequencies into a grid of $N=501$. Experimentally, the numerics can correspond to a pump with a central wavelength of $776$ nm, and degenerate central frequencies of the signal and idler modes of $1552$ nm, and duration  $\sim 200$ fs. The average number of signal photons is also reported to vary between $\langle N_{S} \rangle \approx 3$ to $\langle N_{S} \rangle \approx 10$~~\cite{zhong2020exp1}.
	
	\begin{figure*}[ht]
		\includegraphics[width=\linewidth]{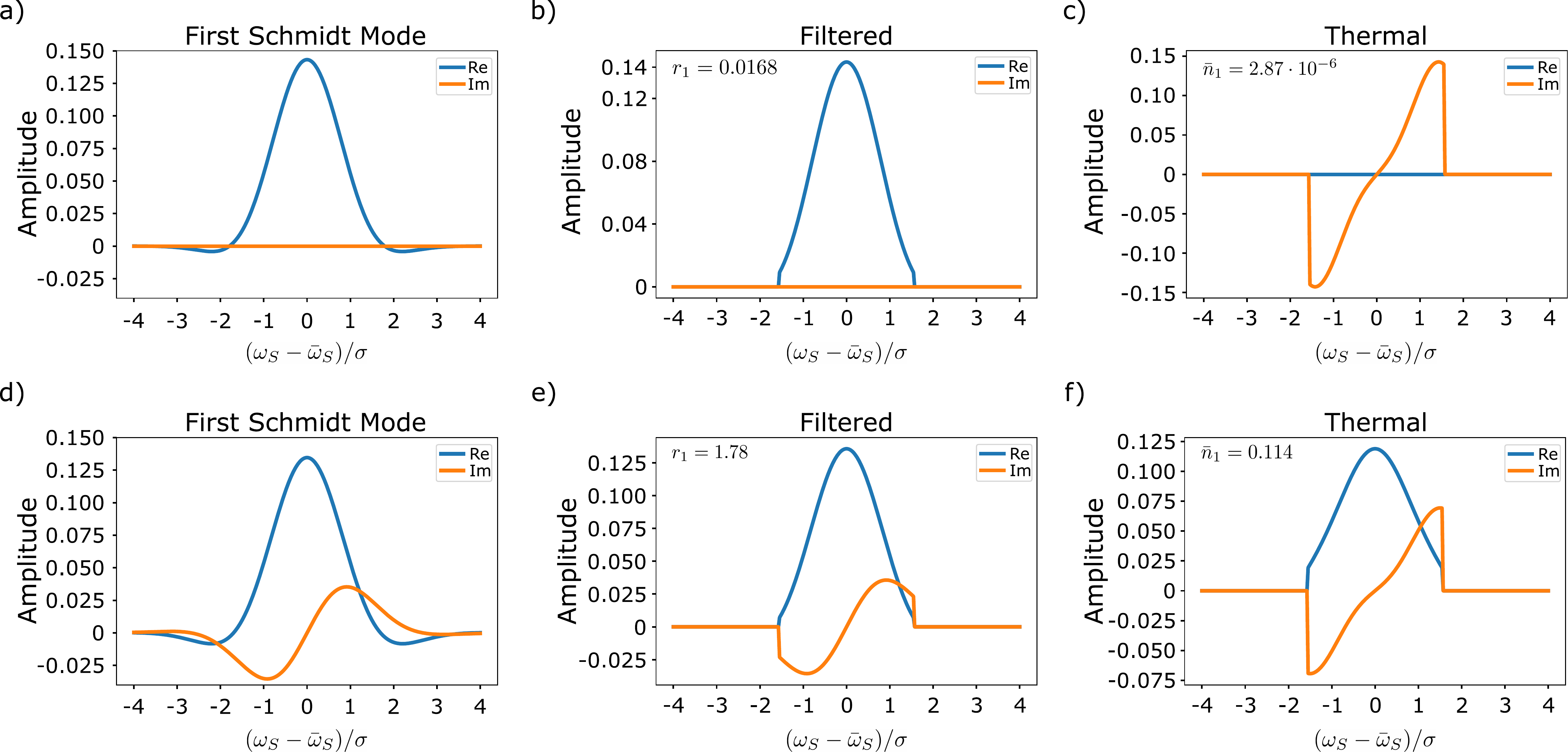}
		\caption{Filtering effects on first Schmidt mode for unapodized nonlinear region with a top-hat filter with range $\left(\omega_{S}-\bar{\omega}_{S}\right)/\sigma \in (-1.5, 1.5)$. a) The first Schmidt mode without filtering in the low-gain regime. b) The first filtered Schmidt mode in the low-gain regime, very similar to just applying the filter to the unfiltered Schmidt mode. c) First thermal mode of the filtered state in the low-gain regime. The thermal mode and filtered Schmidt mode are significantly different. d) The first Schmidt mode without filtering in the high-gain regime. e) The first filtered Schmidt mode in the high-gain regime. Even with filtering low and high gain modes are still highly mismatched. f) First thermal mode of the filtered state in the high-gain regime. The thermal modes in the low and high gain regimes are themselves highly mismatched. Note that in the high-gain, the filtered Schmidt mode and thermal mode are less mismatched. Low(high)-gain regime is taken with $\langle N^{\text{Bare}}_{S} \rangle = 3\cdot10^{-4}(10.6)$. We include the squeezing parameters(thermal occupancies) for the Filtered(Thermal) plots.}
		\label{fig:unpoled}
	\end{figure*}

	\subsection{Joint spectral amplitude and spectral purity}\label{Sec:UPJSA}
	In Fig.~\ref{fig:UnpoledJSA}(a) (Fig.~\ref{fig:UnpoledJSA}(b)) we show the absolute value of the JSA(see Eq.~\ref{eq:JSA}) for the single pass setup in the low(high)-gain regime, with $\langle N_{S} \rangle =3.13\cdot10^{-4}(10.6) $. As expected, in the low-gain regime, the JSA is approximately given by a product of a Gaussian and a Sinc function. Due to the Sinc function, there are many side lobes in the JSA, indicating frequency correlations and thus low spectral purity. In the high-gain regime, we no longer have a product of a Gaussian and a Sinc. Time-ordering corrections give rise to distortions and a broadening of the central portion of the JSA~\cite{quesada2014time,quesada2015time,quesada2022BPP}. In both regimes, there are high levels of frequency correlations which tells us that the squeezed output state is spectrally mixed. As we can see from Fig.~\ref{fig:UnpoledJSA}(e), the Schmidt number decreases as a function of gain. Although the Schmidt number decreases, it remains very well above unity over the range of experimentally relevant gain. In fact, the Schmidt number decreases from $K=1.17$ to $K=1.06$, indicating that the squeezed output state is spectrally mixed for all levels of gain.

	\subsection{Temporal mode structure}
	In Fig.~\ref{fig:UnpoledJSA}(c)(Fig.~\ref{fig:UnpoledJSA}(d)) we show the temporal mode structure of the squeezed output state in the low(high)-gain regime. We plot both the real and imaginary part of the first Schmidt mode $\rho^{S}_{1}(\omega)$(see Eq.~\ref{Eq:Numcor} and subsequent paragraph). Although the Schmidt number is above unity, we only plot the first Schmidt mode for distinguishability purposes. As we increase gain, the temporal modes increasingly pick up an imaginary part that is orthogonal to its real part. This tells us that the squeezed states at different levels of gain are distinguishable. To characterize just how distinguishable the states at different levels of gain are, we consider the fidelity between the frequency profiles of two temporal modes
	\begin{align}\label{eq:fidelity}
		F(A_{i},A_{j}) = \left|\int d\omega A_{i}(\omega)\left(A_{j}(\omega)\right)^{*}    \right|^{2}.
	\end{align}
	We denote $F(L,H)$ to be the fidelity between a fixed low-gain Schmidt mode, $\rho^{S}_{L}(\omega)$, and a variable-gain Schmidt mode, $\rho^{S}_{H}(\omega)$. In Fig.~\ref{fig:UnpoledJSA}(f) we plot $F(L,H)$ as a function of gain in the variable-gain mode. As we increase the gain of the variable-gain Schmidt mode, the fidelity drops to below $90\%$ over the range of experimentally relevant gain. If the squeezed states generated vary greatly in levels of gain, they can be quite distinguishable. This is highly undesirable in the context of GBS where distinguishability can nullify your quantum computational advantage~\cite{Valery2022distinguish,Shi2021distinguish}.

	\begin{figure*}[ht]
		\includegraphics[width=\linewidth]{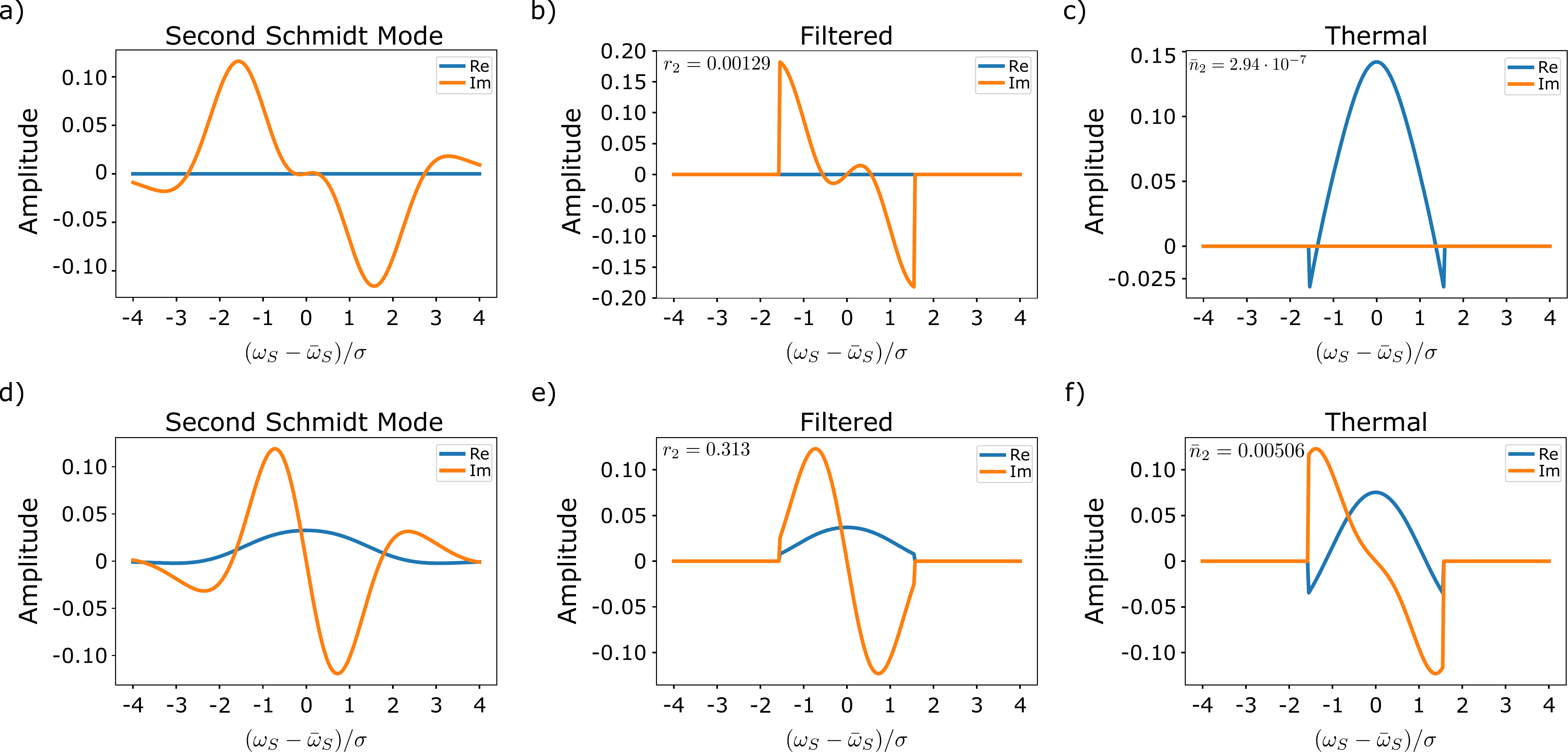}
		\caption{Filtering effects on second Schmidt mode for unapodized nonlinear region with a top-hat filter with range $\left(\omega_{S}-\bar{\omega}_{S}\right)/\sigma \in (-1.5, 1.5)$. a) The second Schmidt mode without filtering in the low-gain regime. b) The second filtered Schmidt mode in the low-gain regime, very similar to just applying the filter to the unfiltered Schmidt mode. c) Second thermal mode of the filtered state in the low-gain regime. The thermal mode and filtered Schmidt mode are significantly different. d) The second Schmidt mode without filtering in the high-gain regime. e) The first filtered Schmidt mode in the high-gain regime. Even with filtering low and high gain modes are still highly mismatched. f) second thermal mode of the filtered state in the high-gain regime. The thermal modes in the low and high gain regimes are themselves highly mismatched. Note that in the high-gain, the filtered Schmidt mode and thermal mode are less mismatched. Low(high)-gain regime is taken with $\langle N^{\text{Bare}}_{S} \rangle = 3\cdot10^{-4}(10.6)$. We include the squeezing parameters(thermal occupancies) for the Filtered(Thermal plots).}
		\label{fig:unpoled2}
	\end{figure*}

	\subsection{Filtering effects}
	We now study how filtering affects the output squeezed state for an unapodized nonlinear region. We consider a top-hat filter with range $\left(\omega_{S}-\bar{\omega}_{S}\right)/\sigma \in (-1.5, 1.5)$. We apply the same filter to both signal and idler modes. Since the output squeezed state is spectrally mixed, we need to describe both the squeezing and the thermal modes. In Fig.~\ref{fig:unpoled}(a) we show the first unfiltered mode as a reference. Fig.~\ref{fig:unpoled}(b) shows the first filtered Schmidt mode. Although we obtain the mode from the $\bm{O}\bm{W}^{T}$ matrix, this mode is almost identical to applying the filter function to the unfiltered mode and renormalizing. Fig.~\ref{fig:unpoled}(c) shows the first thermal mode which differs significantly from the filtered Schmidt mode. In this low-gain limit, filtering allows us to regain spectral purity. By calculating the filtered Schmidt number, we find that it reduces to approximately $1.03$ which is much closer to unity than the unfiltered case.
	
	Although filtering increases the spectral purity, it does not help reduce distinguishability between modes at different levels of gain. The bulk of the distinguishability in the unapodized case comes from the central frequencies which are unaffected by the filter. This is seen by comparing the high-gain modes shown in Fig.~\ref{fig:unpoled}(d)(e)(f) to the low-gain modes. Both the Schmidt and thermal modes are highly mismatched. It is important to note that in the high-gain limit, the filtered Schmidt mode and thermal modes become less mismatched. This can be understood by the fact that as gain increases, the Schmidt number decreases and we get closer to spectral purity. As mentioned in Sec.\ref{sec:pure}, when we are spectrally pure the filtered Schmidt modes and thermal modes are the same. We then expect these modes to become increasingly similar as the Schmidt number approaches unity.
	
	In Fig.~\ref{fig:unpoled2} we show all the same plots but for the second unfiltered Schmidt mode. All the conclusions for the first mode also apply for the second mode. We include this however, to show that filtered and thermal modes of different unfiltered Schmidt modes have non-zero overlap. In the low-gain limit, by comparing Fig.~\ref{fig:unpoled}(b) and Fig.~\ref{fig:unpoled2}(c) we see that the first filtered Schmidt mode and the second thermal mode have similar profiles. The same is true for the second Schmidt mode and the first thermal mode. To illustrate these overlap, we consider a vector of modes $\bm{s}= (A_{1}(\omega),A_{2}(\omega),\mathcal{A}_{1}(\omega),\mathcal{A}_{2}(\omega))^{T}$ where $A_{i}(\omega)$ represents the filtered Schmidt modes and $\mathcal{A}_{i}(\omega)$ the thermal modes, and we consider the fidelity matrix $\mathcal{F}_{i,j}=F(\bm{s}_{i},\bm{s}_{j})$ with the fidelity given by Eq.~\ref{eq:fidelity}. In the low-gain limit we find
	\begin{align}
		\mathcal{F}=\begin{pmatrix}
			1 & 0 & 1.25\cdot 10^{-8} & 0.983 \\
			0 & 1 & 0.879 & 8.18\cdot 10^{-9} \\
			1.25\cdot 10^{-8} & 0.879 & 1 & 0 \\
			0.983 & 8.18\cdot 10^{-9} & 0 & 1 \\
		\end{pmatrix}.
	\end{align}
	As expected, different filtered Schmidt modes are mutually orthogonal and so are the different thermal modes. However, there is non-zero overlap between filtered Schmidt and thermal modes. In the low-gain limit, the overlap between filtered Schmidt and thermal modes is highest between different modes (e.g. between $A_{1}(\omega)$ and $\mathcal{A}_{2}(\omega)$). In the high-gain limit, the overlap matrix is
	\begin{align}
		\mathcal{F}=\begin{pmatrix}
			1 & 0 & 0.930 & 0.0501 \\
			0 & 1 & 6.33\cdot 10^{-3} & 0.563 \\
			0.930 & 6.33\cdot 10^{-3} & 1 & 0 \\
			0.0501 & 0.563 & 0 & 1
		\end{pmatrix}.
	\end{align}
	As we can see, there is still overlap between the filtered Schmidt and thermal modes. However, the dominating overlaps have shifted to be between filtered Schmidt and thermal modes of the same mode. As previously mentioned, in the high-gain limit the Schmidt number approaches unity. We thus expect higher modes to fizzle out and the first filtered Schmidt mode and first thermal mode to become the same.
	
	In both low and high-gain regimes, filtering gives rise to several undesired effects. Although it does help increase spectral purity, the filtered output states contain both a squeezed and thermal portion which are distinguishable between modes. We conclude that an unapodized nonlinear region is not an ideal source for squeezed state generation both with and without filtering.
 
	\begin{figure*}[ht]
		\includegraphics[width=\linewidth]{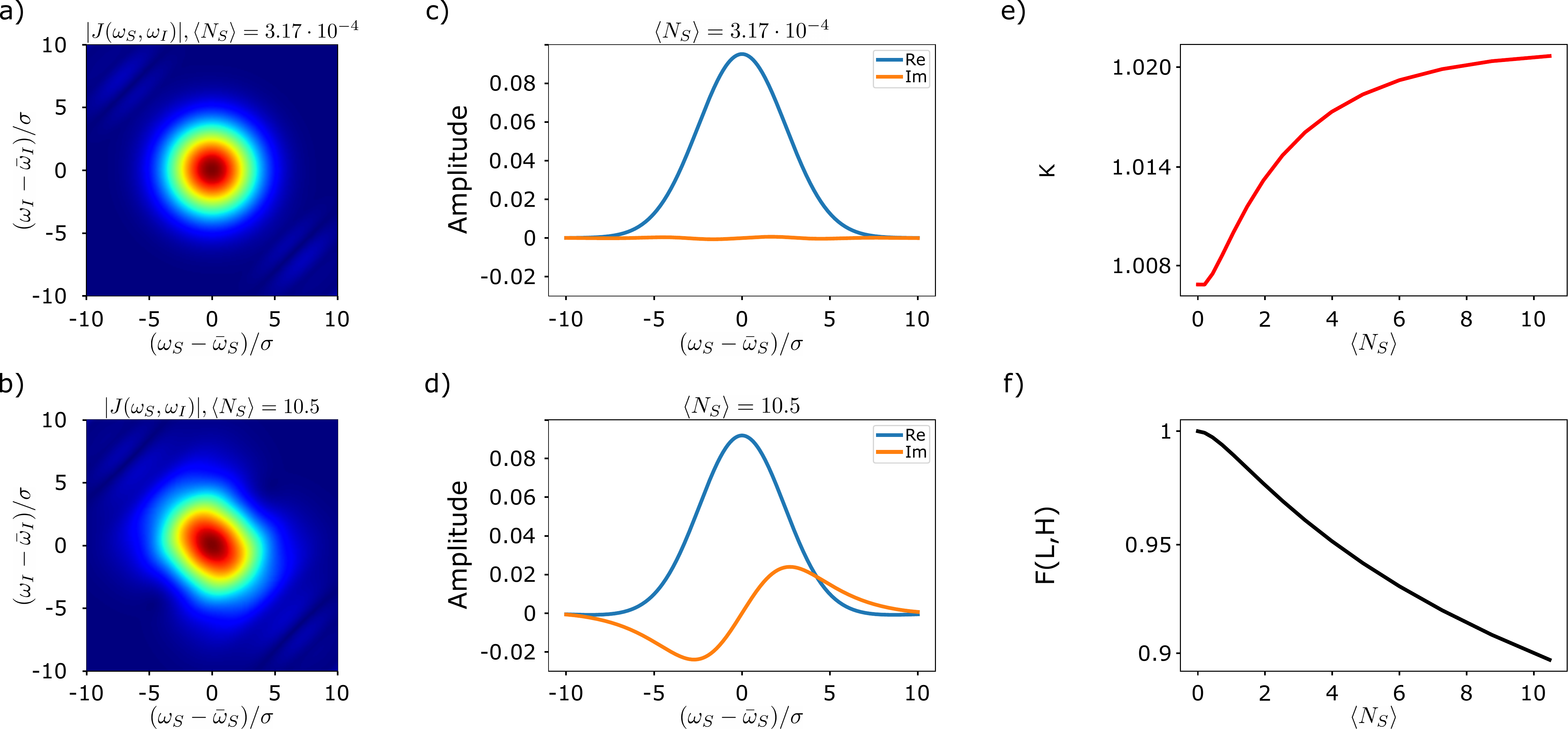}
		\caption{Results for the apodized single pass setup (c.f. Sec.~\ref{sec:SPR}). a) Absolute value of the JSA is the low-gain regime with $\langle N_{S} \rangle = 3.17\cdot 10^{-4}$. As expected by the choice of poling function, the JSA is approximately a product of two Gaussians. b) Absolute value of the JSA in the high-gain regime with $\langle N_{S} \rangle = 10.5$. High-gain effects leads to a warping of the Gaussians and a decrease in spectral purity. c) First Schmidt mode, $\rho^{S}_{1}(\omega)$, in the low-gain regime with $\langle N_{S} \rangle = 3.17\cdot 10^{-4}$. d)  First Schmidt mode, $\rho^{S}_{1}(\omega)$, in the high-gain regime with $\langle N_{S} \rangle = 10.5$. In the high-gain regime, the first Schmidt mode has a much greater imaginary part telling us that the Schmidt modes at different gains are distinguishable. e) The Schmidt Number, $K$, as a function of gain, characterized by $\langle N_{S} \rangle$. As gain increases, $K$, increases which indicates a loss of spectral purity. However, $K$ remains relatively close to unity. f) The fidelity, $F(L,H)$, between the first Schmidt mode at fixed low-gain, $\rho^{S}_{L}(\omega)$, and the first Schmidt mode at variable-gain, $\rho^{S}_{H}(\omega)$, as a function of gain. The fidelity drops to slightly below $90\%$ which indicates partial distinguishability of Schmidt modes at different levels of gain.}
		\label{fig:Single}
	\end{figure*}

	\section{Apodized Single Pass Results}
	\label{sec:SPR}

	We now consider the squeezed output states generated by the setup of Fig.~\ref{fig:model}(b) with the nonlinear region poled such that in the low-gain regime, the phase-matching function is given by a Gaussian. As in the unapodized case, we consider the nonlinear region to be of length $l$ and allow for type-\RomanNumeralCaps{2} SPDC processes. To mimic such a poling function, we follow the methods of ~\cite{tambasco2016domain,Graffitti_2017,Aggie2021domain}, using the Python library \texttt{Custom-Poling}~\cite{custompoling} we break the nonlinear region into $N_{z} = 1000$ domains. Optimizing gives us the proper signs for $g(z)$ to obtain an approximate Gaussian phase-matching function(see Appendix~\ref{app:poling} for more information on $g(z)$ and the width of the Gaussian phase-matching function).
	
	We study the squeezed output states of the poled single pass structure by taking the pump mode to be the same as in the unapodized case (see Eq.~\ref{Eq:pumpform}) and again assume that we are in the symmetric group-velocity-matched regime (see Eq.~\ref{Eq:symgrpvel}). We solve numerically for the propagator by breaking the frequencies into a grid of $N=501$. All experimental properties are the same as those mentioned in Sec.~\ref{sec:Unpoled}.
	
	\begin{figure*}[ht]
		\includegraphics[width=\linewidth]{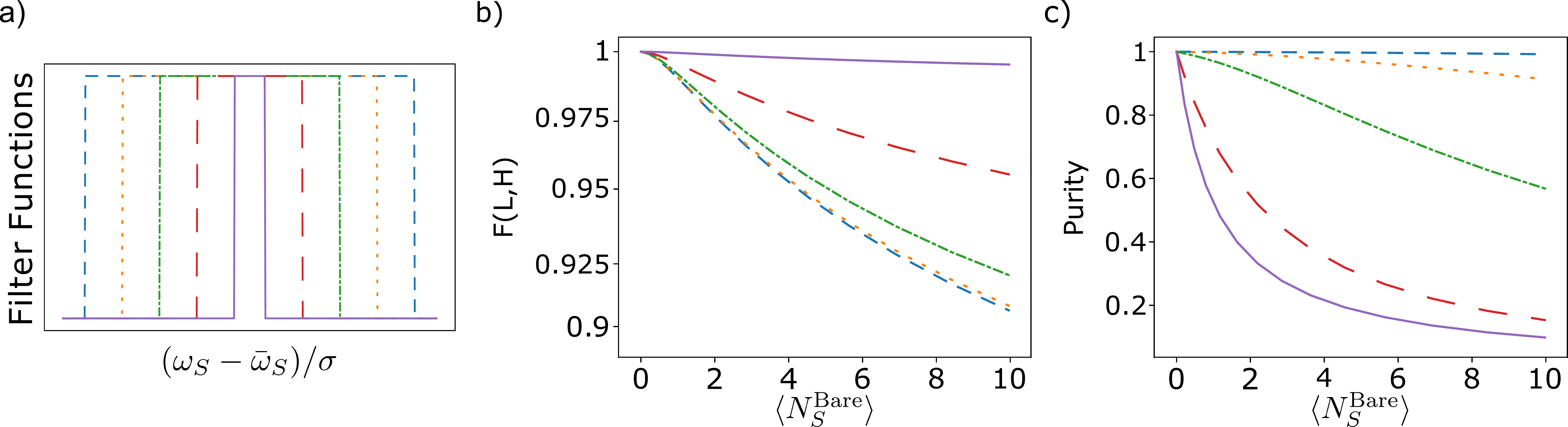}
		\caption{Filtering and its effect on fidelity and purity for apodized single pass. a) Different filter functions used. b)The fidelity, $F(L,H)$, between the first Schmidt mode at fixed low gain, $\rho^{S}_{L}(\omega)$, and the first Schmidt mode at variable gain, $\rho^{S}_{H}(\omega)$, as a function of of the pre-filtered gain, $\langle N^{\text{Bare}}_{S}\rangle$ (e.g. the number of signal photons observed before filtering). As the filter width narrows, the fidelity increased for all levels of gains. c) Purity of the filtered squeezed state as a function of pre-filtered gain. Note that this is not the same as the spectral purity. The narrower the filter width, the more detrimental the effects to purity as a function of gain are. Although filtering increases fidelity, it comes at the cost of decreasing purity. The colours and markers of the filter functions correspond to the different curves in b) and c).}
		\label{fig:Filter}
	\end{figure*}

	\subsection{Joint spectral amplitude and spectral purity}\label{Sec:SPRJSA}
	In Fig.~\ref{fig:Single}(a) (Fig.~\ref{fig:Single}(b)) we show the absolute value of the JSA for the single pass setup in the low(high)-gain regime, with $\langle N_{S} \rangle =3.17\cdot10^{-4}(10.5) $. As expected, in the low-gain regime, the JSA is approximately given by a product of two Gaussians. The side fluctuations that we see are caused by the fact that the poling function, $g(z)$, only centrally approximates a Gaussian and has some small non-vanishing tails. In the high-gain regime, we no longer have a product of two Gaussians. Time-ordering corrections give rise to distortions and a broadening of the JSA~\cite{quesada2014time,quesada2015time,quesada2022BPP}. High-gain effects lead to an increase in the frequency correlations of the squeezed state and thus a decrease in spectral purity. As we can see from Fig.~\ref{fig:Single}(e), the Schmidt number increases as a function of gain. For the range of gain that is experimentally relevant, the variations are quite small. In fact, the Schmidt number increases from $K=1.0064$ to $K=1.0188$ which is still fairly low. This tells us that even in the high-gain regime, the JSA remains separable to a good approximation. Although we remain spectrally pure, the temporal mode structure tells a different story.
	
	\subsection{Temporal mode structure}
	In Fig.~\ref{fig:Single}(c)(Fig.~\ref{fig:Single}(d)) we show the temporal mode structure of the squeezed output state in the low(high)-gain regime. As the Schmidt number remains relatively low, we only plot the first Schmidt mode (the only relevant one). As in the unapodized case, when we increase gain, the temporal modes increasingly pick up an imaginary part that is orthogonal to its real part. The squeezed states at different levels of gain for the single pass case are thus also distinguishable.  In Fig.~\ref{fig:Single}(f) we plot the fidelity between a fixed low-gain Schmidt mode and a variable-gain Schmidt mode. Similarly to the unapodized case, the fidelity drops to below $90\%$ over the range of experimentally relevant gain. Again, this is highly undesirable in the context of GBS where distinguishability nullifies your quantum computational advantage~\cite{Valery2022distinguish,Shi2021distinguish}. Variations in pump intensities still gives rise to partially distinguishable squeezed states in an apodized single pass setup.

	\subsection{Fidelity and Purity}
	We now study how filtering affects the fidelity between states with different levels of gain as well as the purity of the output state. From the way the Schmidt modes vary as a function of frequency (see Fig.~\ref{fig:Single}(c)(d)), one can hope that removing the tail ends of the modes will help decrease distinguishability. To see if this is the case, we consider how filtering affects the squeezed output state.
	
	As mentioned in Sec.~\ref{Sec:SPRJSA}, the Schmidt number of the squeezed output states remains relatively close to unity. We thus approximate the state to be described by a singular Schmidt mode and obtain the temporal mode structure of the filtered state as prescribed in Sec.~\ref{sec:pure}.
	
	We consider several different top-hat filter functions of varying width as shown in Fig.~\ref{fig:Filter}(a). Each filter is applied to both signal and idler modes. Fig.~\ref{fig:Filter}(b) shows the fidelity (Eq.~\ref{eq:fidelity})
	between a fixed low-gain filtered Schmidt mode and a variable-gain filtered Schmidt mode. Since each different filter width renormalizes the average number of signal photons differently, we characterize the level of gain by the pre-filtered gain which we label by $\langle N^{\text{Bare}}_{S}\rangle$(given by Eq.~\ref{Eq:Ns}). As the filter width narrows, the fidelity increases for all values of gain. Filtering does in fact decrease distinguishability and the narrower the filter width, the better the decrease. This, however, comes with a price.
	
	Fig.~\ref{fig:Filter}(c) shows the purity of the filtered state, given by $P=\text{tr}(\varrho^{2}_{th-SS})$ as a function of the pre-filtered gain. Note that this is not the same as the spectral purity. For pure states, we have that $P=1$ whereas for mixed states $P<1$. When using squeezed states as resources, one always wants the state to be as pure as possible~\cite{braunstein2005info}. As we can see, purity decreases as a function of gain for all different filter widths. It decreases even more rapidly the narrower the filter is. Hence, although a narrower filter decreases distinguishability more, it comes at the price of increasingly decreasing the purity of the state. Filtering gives rise to a fundamental trade-off between distinguishability and purity.

	\begin{figure*}[t]
		\includegraphics[width=\linewidth]{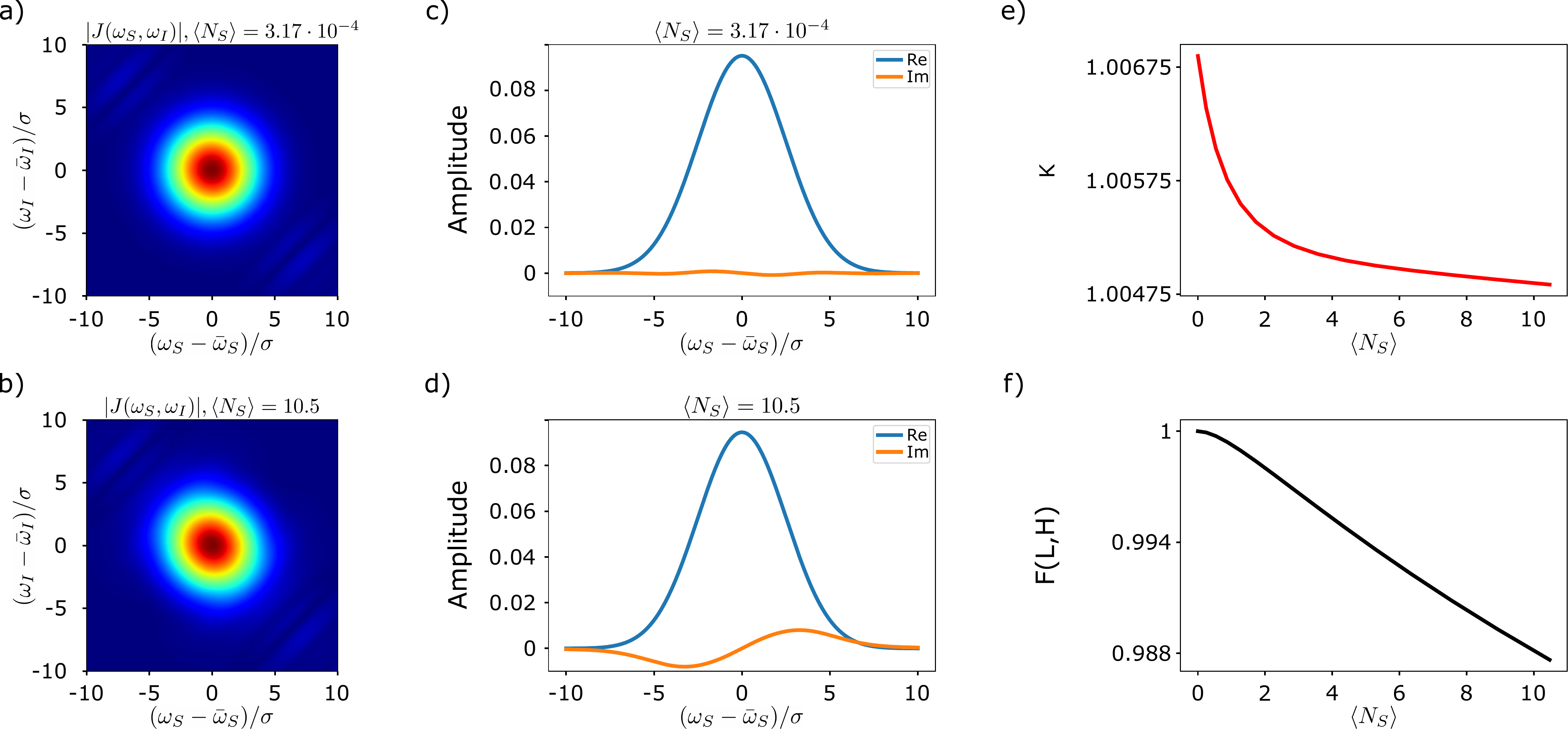}
		\caption{Results for the double pass setup (c.f. Sec.~\ref{sec:DP}). a) Absolute value of the JSA is the low-gain regime with $\langle N_{S} \rangle = 3.17\cdot 10^{-4}$. As expected by the choice of poling function, the JSA is approximately a product of two Gaussians. b) Absolute value of the JSA in the high-gain regime with $\langle N_{S} \rangle = 10.5$. High-gain effects are much less noticeable than in the single pass case. c) First Schmidt mode, $\rho^{S}_{1}(\omega)$, in the low-gain regime with $\langle N_{S} \rangle = 3.17\cdot 10^{-4}$. d)  First Schmidt mode, $\rho^{S}_{1}(\omega)$, in the high-gain regime with $\langle N_{S} \rangle = 10.5$. In the high-gain regime, the imaginary part is much less pronounced than in the single pass case and thus the modes are less distinguishable. e) The Schmidt Number, $K$, as a function of gain, characterized by $\langle N_{S} \rangle$. As gain increases, $K$, decrease which indicates a gain in spectral purity for the double pass case. f) The fidelity, $F(L,H)$, between the first Schmidt mode at fixed low-gain, $\rho^{S}_{L}(\omega)$, and the first Schmidt mode at variable-gain, $\rho^{S}_{H}(\omega)$, as a function of gain. The fidelity remains much higher as a function of gain in the double pass case, indicating that Schmidt modes at different levels of gain are more indistinguishable.}
		\label{fig:Double}
	\end{figure*}
	
	\section{Apodized Double Pass Results}
	\label{sec:DP}
	We now consider the squeezed output states generated by a double pass structure shown in Fig.~\ref{fig:model}(c). As in the previous setups, we consider the nonlinear regions to allow for type-\RomanNumeralCaps{2} SPDC processes.
	
	In this setup, the pump is sent through a first nonlinear region and generates pairs of signal and idler modes of opposite polarization. The three modes then propagate freely in vacuum and pass through a half-wave plate(HWP). This HWP does not affect the pump, however, it swaps the polarizations of the signal and idler modes. At the level of the equations of motion (Eqs.~\ref{Eq:SigEomZ},\ref{Eq:IdEomZ}) this simply interchanges $v_{S}\leftrightarrow v_{I}$. The modes then propagate freely again until they pass through a second nonlinear region.   
	
	As in the apodized single pass case, we take the first nonlinear region to be poled such that we obtain an approximate Gaussian phase-matching function in the low-gain regime. To mimic a setup where a mirror is used to reflect the modes back towards the nonlinear region, we flip the poling of the second nonlinear region. Both nonlinear regions have the same length $l$ and are broken into $N_{z}=1000$ domains. Furthermore, when solving for the propagator of the second nonlinear region, we swap the signal and idler velocities to mimic the polarization swap.  
	
	We know how to obtain the propagators for both nonlinear regions, but now we need to consider the free propagation in between regions. Assuming vacuum, all three modes propagate at the same the velocity, $c$ the speed of light, in between the nonlinear regions. Working with the Fourier transformed operators (Eq.~\ref{Eq:Fouriermode}) that move at the pump velocity, we find that in free space the equations of motion reduce to
	\begin{align}
		\frac{\partial}{\partial z} a_{j}(z,\omega)=0,
	\end{align}
	which gives the trivial propagator $\bm{U}(z,z_{0}) = \mathbb{1}$. Assuming that reflections at any of the interfaces are negligible, the full solution for the propagator is simply given by combining the propagator for each individual region (e.g. $\bm{U}_{\text{Total}} = \bm{U}_{2}\mathbb{1}\bm{U}_{1}$).
	
	We study the squeezed output states of the double pass structure by taking the pump mode to be the same as in the unapodized case (see Eq.~\ref{Eq:pumpform}) and again assume that we are in the symmetric group-velocity-matched regime (see Eq.~\ref{Eq:symgrpvel}). We solve numerically for the propagator for each region by breaking the frequencies into a grid of $N=501$. All experimental properties are the same as those mentioned in Sec.~\ref{sec:Unpoled}.
	
	\subsection{Joint spectral amplitude and spectral purity}
	In Fig.~\ref{fig:Double}(a)(Fig.~\ref{fig:Double}(b)) we show the absolute value of the JSA for the double pass setup in the low(high)-gain regime, with $\langle N_{S} \rangle = 3.17\cdot 10^{-4}(10.5)$. In the low-gain regime, the double pass setup gives almost the same absolute JSA as the single pass case. In both cases, we obtain a JSA that is approximated by a product of two Gaussians. Again, the side fluctuations are due to the poling function only centrally approximating a Gaussian phase-matching function. In the high-gain regime, we see a different result. The high-gain effects are less pronounced and the JSA becomes only slightly more elliptical along one of its axes. By comparing Fig.~\ref{fig:Single}(b) and Fig.~\ref{fig:Double}(b), we see that the double pass setup leads to less detrimental high-gain effects and thus a higher spectral purity than the single pass setup. In Fig.~\ref{fig:Double}(e) we plot the Schmidt number (Eq.~\ref{Eq:schmidt}) as a function of gain. As gain increases, the Schmidt number slightly decreases. This tells us that spectral purity increases in the double pass setup and that the JSA remains separable to an even better approximation in the high-gain regime.   
	
	\subsection{Temporal mode structure}
	In Fig.~\ref{fig:Double}(c)(Fig.~\ref{fig:Double}(d)) we show the temporal mode structure of the squeezed output state in the low(high)-gain regime. We again only show the first Schmidt mode since we are to a good approximation spectrally pure. As we increase the gain, the temporal modes still acquire an imaginary part that is orthogonal to its real part. However, this imaginary part is much less pronounced than in the single pass case. Although the squeezed output states at different levels of gain are still slightly distinguishable, they are much less so than in the single pass case. In Fig.~\ref{fig:Double}(f) we show the fidelity (Eq.~\ref{eq:fidelity}) between a fixed low-gain Schmidt mode and a variable-gain Schmidt mode as a function of gain. Over the same range of gain, the fidelity now only drops to below $98.8\%$ which is a much better result than in the single pass case. As shown by the fidelity, the squeezed output states with different levels of gain are much less distinguishable in the double pass case and do not require to be additionally filtered.  
	
	\subsection{Comparison}
	By comparing results of the two different apodized geometries, we see that the double pass setup substantially gives better results when considering both the spectral purity and distinguishability. We can heuristically understand why this is so by considering the effects of the polarization swap on the propagation through the second nonlinear region. In the symmetric group-velocity matched regime the polarization swap leads to the $\Delta k_{l}(\omega)$ terms in Eqs.~\ref{Eq:SigEomZ} and \ref{Eq:IdEomZ} to change signs. When the signal and idler modes pass through the second nonlinear region, they are then partially subject to a different and opposite evolution. Thus, as they pass through the second nonlinear region, part of the evolution from the first pass is cancelled out, leading to a smaller imaginary part in the temporal mode structure. In accordance with the authors' claims and the statistical analysis of Ref.~~\onlinecite{martinez2022classical}, we have shown that the sources used in Ref.~~\onlinecite{zhong2021exp2} are better suited for quantum sampling than those used in Ref.~~\onlinecite{zhong2020exp1}.

	\section{Conclusion}
	\label{sec:conc}
	
	In this paper, we have proposed a physical mechanism to explain partial distinguishability between squeezed states generated by a parametric waveguided source driven by pumps with different brightness but identical profiles. We considered a single pass geometry with both unapodized and apodized nonlinear regions as well as a double pass geometry with apodized nonlinear regions. In all setups, we study the temporal mode structure of the squeezed output state as a function of gain, which is only modified through the number of pump photons $\langle N_{P} \rangle$.
	
	In the unapodized single pass setup, we find that the squeezed output state is spectrally mixed for all levels of gain. Although filtering can be used to increase spectral purity, we never obtain a separable JSA. The filtered output state can not be approximated by one Schmidt mode. Furthermore, the filtered output state is described by both a squeezing and thermal part which are distinguishable both for a single mode and between different modes.
	
	In the apodized single pass setup, we find that the squeezed output state remains, to a good approximation, spectrally pure as a function of gain. However, we find that states with different levels of gain can be quite distinguishable with fidelities going below $90\%$. This is highly relevant to GBS where distinguishability can completely negate any quantum computational advantage. By studying common filtering techniques used to decrease distinguishability, we find that although we do achieve better distinguishability after filtering, it comes at the cost of decreasing the purity of the squeezed output state. This is an important result as we typically require squeezed states to be as pure as possible to maximize its use as a resource for quantum processes.    
	
	In the double pass setup, we find that the output squeezed state remains, to an even better approximation, spectrally pure as a function of gain. Due to the polarization swap in between the two nonlinear regions, the squeezed output states with different levels of gain are far less distinguishable than the single pass setup. In fact, in this setup, the fidelity only drops to $98.8\%$. This value is above the threshold for which GBS can show quantum computational advantage ~\cite{Shi2021distinguish}. Since the distinguishability is low, we do not study filtering techniques for this setup. Notably, the double pass structure fulfills the criteria required for practically useful continuous variable quantum computing outlined in ~\cite{vernon2019scalable}.
	
	The results presented are relevant to any setup that uses similar squeezed output states~\cite{zhong2020exp1,zhong2021exp2,Reddy2017conversion, Reddy2017ramsey}. Our work represents the first exploration of the effects of varying pump intensities on the distinguishability of generated squeezed states in parametric waveguided sources.
	
	Having built a framework to study  double pass configurations, an interesting question to consider next is whether configurations of more passes with more interesting geometries can lead to better results or even topological effects.
	
	\section*{Acknowledgements}
	The authors acknowledge support from the Minist\`{e}re de l'\'{E}conomie et de l’Innovation du Qu\`{e}bec and the Natural Sciences and Engineering Research Council of Canada. The authors thank M. Walschaers, F. Arzani, J. Pereira, and W. McCutcheon for insightful discussions concerning Bloch-Messiah decomposition and its implementation in Python, J.E. Sipe for insightful discussions and W. McCutcheon and L. G. Helt for a critical reading of the manuscript.
	
   \section*{Author declarations}
   \subsection*{Conflict of Interests}
    The authors have no conflicts to disclose.

    \subsection*{Author Contribution}
    \textbf{Martin Houde}: Conceptualization (equal), Formal analysis (lead), Investigation (lead), Methodology (equal), Resources (equal), Software (lead), Validation (equal), Visualization (lead), Writing – original draft (lead), Writing – review \& editing (lead). \textbf{Nicol\'as Quesada}: Conceptualization (lead), Formal analysis (equal), Funding acquisition (lead), Methodology (equal), Project administration (lead), Resources (equal), Software (supporting), Supervision (lead), Validation (equal), Visualization (supporting), Writing – review \& editing (equal).

    \subsection*{Supporting Data}
    The data that supports the findings of this study are available within the article and its supplementary material.

	\appendix
	
	\section{Detailed derivation}\label{app:derivation}
	The canonical formalism developed in ~\cite{quesada2020theory,quesada2022BPP} includes modal and material dispersion and gives rise to Heisenberg equation of motions for the displacement ($\bm{D}(\bm{r})$) and magnetic ($\bm{B}(\bm{r})$) operator fields which replicate the dynamical Maxwell equations. Without loss of generality, we assume that the waveguide propagation is in the positive $z$ direction and we break the Hamiltonian into linear and nonlinear parts such that
	\begin{align}\label{Eq:Hfull}
		H = H_{L} +\sum_{n=2}H^{(n)}_{NL},
	\end{align}
	where
	\begin{align}\label{Eq:Hlin}
		H_{L} = \int dk \sum_{j} \hbar \omega_{j,k}b^{\dagger}_{j,k}b_{j,k},
	\end{align}
	and $H^{(n)}_{NL}$ represents increasingly nonlinear terms coming from expanding the polarization. The index $j$ labels the waveguide modes which obey canonical bosonic commutation relations
	\begin{align}
		[b_{j,k},b^{\dagger}_{j',k'}]=\delta_{j,j'}\delta(k-k'),\\
		[b_{j,k},b_{j',k'}]=[b^{\dagger}_{j,k},b^{\dagger}_{j',k'}]=0.
	\end{align}
	
	As we are only interested in the effects of initially varying pump intensities in SPDC processes, both ignoring self and cross-phase modulations, we only focus on the $n=2$ nonlinear interaction given by
	\begin{align}\label{Eq:Hnlin}
		H^{(2)}_{NL} = -\frac{1}{3\epsilon_{0}}\int d^{3}\bm{r} \Gamma^{(2)}_{j,l,m}(\bm{r})D^{j}(\bm{r})D^{l}(\bm{r})D^{m}(\bm{r}),
	\end{align}
	where we use the Einstein summation convention over Cartesian indices. The $\Gamma$ tensor is linearly related to the susceptibility tensor $\chi^{(2)}_{j,l,m}$ which characterizes the nonlinear response via
	\begin{align}\label{Eq:Gamma}
		\Gamma^{(2)}_{j,l,m}(x,y,z) = \frac{\chi^{(2)}_{j,l,m}(x,y,z)}{\epsilon_{0}n_{0}^{6}(x,y)},
	\end{align}
	where the effects of material dispersion are neglected for the nonlinear Hamiltonian and $n_{0}(x,y)$ is the index of refraction evaluated at some central frequency of interest~\cite{volkov2004nonlin}.
	
	For each relevant waveguide mode, we associate a central wavevector $\bar{k}_{j}$ with a central frequency $\bar{\omega}_{j} = \omega_{j,\bar{k}_{j}}$ and expand
	\begin{align}\label{Eq:Dispersion}
		\omega_{j,k} \approx \bar{\omega}_{j} +v_{j}(k-\bar{k}_{j}),
	\end{align}
	where the group velocity $v_{j}$ is taken to be constant over the frequency ranges of interest and we ignore group velocity dispersions. We then introduce Fourier transformed operators
	\begin{align}\label{Eq:Psidef}
		\psi_{j}(z) = \int \frac{dk}{\sqrt{2\pi}} e^{i(k-\bar{k}_{j})z}b_{j,k},
	\end{align}
	which are centered at $\bar{k}_{j}$ and slowly varying in space. Assuming that the modes of interest span different wavevector and frequency ranges (which is the case for type-\RomanNumeralCaps{2} SPDC processes), we can formally extend the bounds of integration to $\pm \infty$ which in turn makes these new operators also obey canonical bosonic commutation relations
	\begin{align}
		[\psi_{j}(z),\psi^{\dagger}_{j'}(z')]=\delta_{j,j'}\delta(z-z'),\\
		[\psi_{j}(z),\psi_{j'}(z')]=[\psi^{\dagger}_{j}(z),\psi^{\dagger}_{j'}(z')]=0.
	\end{align}
	Substituting the above, we can express the linear Hamiltonian of Eq.~\ref{Eq:Hlin} as
	\begin{align}\label{Eq:HlinPsi}
		H_{L} =& \int dz \sum_{j}\hbar \bar{\omega}_{j}\psi^{\dagger}_{j}(z)\psi_{j}(z)\nonumber\\
		&+\int dz \sum_{j}\frac{i\hbar v_{j}}{2}\left(\frac{\partial \psi^{\dagger}_{j}(z)}{\partial z}\psi_{j}(z) -\psi^{\dagger}_{j}(z)\frac{\partial \psi_{j}(z)}{\partial z}     \right).
	\end{align}
	Furthermore, the Fourier transformed modes allow us to approximate the $\bm{D}(\bm{r})$ field and obtain an expression for the nonlinear Hamiltonian. In general, we can express the displacement field as an expansion of waveguide modes $j$ in the transverse plane $\bm{d}_{j,k}(x,y)$ as
	\begin{align}\label{Eq:Dfull}
		\bm{D}(\bm{r}) = \int \frac{dk}{\sqrt{2\pi}}\sum_{j}\sqrt{\frac{\hbar \omega_{j,k}}{2}}\bm{d}_{j,k}(x,y)e^{ikz}b_{j,k} +\text{H.c.}.
	\end{align}
	By expanding around the central wavevectors $\bar{k}_{j}$, assuming that variations in $\bm{d}_{j,k}(x,y)$ are negligible due to the weakness of the nonlinearity, and substituting Eq.~\ref{Eq:Psidef} we can write the displacement field as
	\begin{align}\label{Eq:Dapprox}
		\bm{D}(\bm{r})\approx \sum_{j}\sqrt{\frac{\hbar \bar{\omega}_{j,\bar{k}_{j}}}{2}}\bm{d}_{j,\bar{k}_{j}}(x,y)\psi_{j}(z)+\text{H.c.},
	\end{align}
	which allows us to obtain a useful expression for the nonlinear Hamiltonian.
	
	We now focus on type-\RomanNumeralCaps{2} SPDC processes where a pump mode is converted into signal and idler modes of different polarizations but identical frequencies. We label the modes of interest with $j=P,S,I$ for pump, signal, and idler mode. For such processes to occur, we demand energy and momentum conservation requiring that
	\begin{align}
		\bar{\omega}_{P}-\bar{\omega}_{S}-\bar{\omega}_{I}&=0,\\
		\bar{k}_{P}-\bar{k}_{S}-\bar{k}_{I}&=0.\label{Eq:Momcon}
	\end{align}
	Note that if quasi-phase matching is used, the right-hand side of Eq.~\ref{Eq:Momcon} should be changed to $\pm 2\pi / \Lambda_{\text{pol}}$ where $\Lambda_{\text{pol}}$ is the poling period.
	
	Under these conditions, the nonlinear Hamiltonian of Eq.~\ref{Eq:Hnlin} takes the form
	\begin{align}\label{Eq:HnlinPsi}
		H^{(2)}_{NL} = -\hbar \int dz \xi(z) \psi^{\dagger}_{S}(z)\psi^{\dagger}_{I}(z)\psi_{P}(z) +\text{H.c.}
	\end{align}
	where
	\begin{widetext}
		\begin{align}
			\xi(z) &= \frac{2}{\epsilon_{0}\hbar}\sqrt{\frac{\hbar^{3}\bar{\omega}_{S}\bar{\omega}_{I}\bar{\omega}_{P}}{2^3}}\int dxdy \Gamma^{(2)}_{j,l,m}(\bm{r})\left[d^{j}_{S,k_{S}}(x,y)\right]^{*}\left[d^{l}_{I,k_{I}}(x,y)\right]^{*}\left[d^{m}_{P,k_{P}}(x,y)\right]\nonumber\\
			&={\epsilon_{0}\hbar}\sqrt{\frac{\hbar^{3}\bar{\omega}_{S}\bar{\omega}_{I}\bar{\omega}_{P}}{2^3}}\int dxdy \frac{\chi^{2}_{j,l,m}(\bm{r})}{\epsilon_{0}n^{2}_{j}n^{2}_{l}n^{2}_{m}(x,y)}\left[d^{j}_{S,k_{S}}(x,y)\right]^{*}\left[d^{l}_{I,k_{I}}(x,y)\right]^{*}\left[d^{m}_{P,k_{P}}(x,y)\right],
		\end{align}
	\end{widetext}
	is the nonlinear coupling parameter for this process. Note that this coupling is only non-zero in the nonlinear region and the $z$-dependence reflects the possible poling configuration. Also note that for SPDC processes, third-order nonlinear terms are also energy(momentum) conserving and give rise to self and cross-phase modulation terms which are being neglected here.
	
	With a full expression for the Hamiltonian we now move our attention to the dynamics of the modes. We treat the mode operators as Heisenberg operators and study their dynamics by deriving their Heisenberg equation of motions given by
	\begin{align}\label{Eq:Heisenberg}
		i\hbar \frac{d}{dt}O(t) = [O(t),H_{L}+H^{(2)}_{NL}]
	\end{align}
	for an arbitrary operator $O$.
	
	\subsection{Pump dynamics}
	We begin by considering the Heisenberg equation of motion for the pump mode. Substituting Eqs.~\ref{Eq:HlinPsi} and \ref{Eq:HnlinPsi} in Eq.~\ref{Eq:Heisenberg} and setting $O(t)=\psi_{P}(z,t)$ we find
	\begin{align}\label{Eq:PumpEom}
		\left(\frac{\partial}{\partial t}+v_{P}\frac{\partial}{\partial z}+i\bar{\omega}_{P}\right)\psi_{P}(z,t) = i\xi^{*}(z)\psi_{S}(z,t)\psi_{I}(z,t),
	\end{align}
	where the right-hand side is known as a "back-action" term. As is usual, we assume that the pump mode is prepared in a strong coherent state with a large number of photons which remains constant throughout the interaction (undepleted-classical pump approximation). Under these assumptions, we can ignore the back-action term and we replace $\psi_{P}(z,t)$ by its mean field value $\langle \psi_{P}(z,t)\rangle$. The solution is then given by
	\begin{align}\label{Eq:Pumpsoln}
		\langle \psi_{P}(z,t)\rangle = \Lambda[z-v_{P}(t-t_{0})]e^{-i\bar{\omega}_{P}(t-t_{0})},
	\end{align}
	where $\langle \psi_{P}(z,t_{0})\rangle = \Lambda[z]$ is the spatial pump envelope function which is normalized according to the mean number of pump photons given by
	\begin{align}
		N_{P} = \int dz | \langle\psi_{P}(z,t)\rangle |^{2}=\int dz |\Lambda[z]|^{2}\gg 1.
	\end{align}
	
	\subsection{Twin-beam dynamics}
	Having obtained a solution for the mean field value of the pump mode, we can now calculate the Heisenberg equations of motion for the signal and idler modes
	\begin{align}
		\left(\frac{\partial}{\partial t}+v_{S}\frac{\partial}{\partial z}+i\bar{\omega}_{S}\right)\psi_{S}(z,t) =i\xi(z)\langle \psi_{P}(z,t)\rangle\psi^{\dagger}_{I}(z,t),\label{Eq:SigEom} \\
		\left(\frac{\partial}{\partial t}+v_{I}\frac{\partial}{\partial z}-i\bar{\omega}_{I}\right)\psi^{\dagger}_{I}(z,t) = i\xi^{*}(z)\langle \psi^{\dagger}_{P}(z,t)\rangle\psi_{S}(z,t).\label{Eq:IdEom}
	\end{align}
	As was expected, the right-hand side of Eqs.~\ref{Eq:SigEom} and \ref{Eq:IdEom} only accounts for photon generation via SPDC where as the left-hand side takes account for the propagation at group velocity $v_{j}$, and oscillations at frequency $\bar{\omega}_{j}$.
	
	To bring the equations of motion to a form that is more readily solvable numerically, we introduce new operators for the the signal and idler fields
	\begin{align}
		a_{l}(z,\omega) &= \int \frac{dt}{\sqrt{2\pi/v_{j}}}e^{i(\omega t-z(\omega -\omega_{l})/v_{P})}\psi_{l}(z,t)\label{Eq:Fouriermode},\\
		\psi_{l}(z,t)&=\int \frac{d\omega}{\sqrt{2\pi v_{l}}}e^{-i(\omega t-z(\omega -\omega_{l})/v_{P})}a_{l}(z,\omega),
	\end{align}
	where the index $l=S,I$ exclusively. Note that these new operators are simply the time-to-frequency Fourier transforms of the $\psi_{j}$ operators in a moving frame at the group velocity of the pump, $v_{P}$~\cite{Vidrighin2017QuantumOM}. As shown in ~\cite{quesada2020theory}, these new operators obey canonical bosonic commutation relations for all $z$ and we can interpret quantities such as $a^{\dagger}_{l}(z,\omega)a_{l}(z,\omega)$ as a photon frequency density operator at position $z$. Substituting the new operators in the equations of motion leads to Eqs.~\ref{Eq:SigEomZ} and \ref{Eq:IdEomZ} of the main text.

	\begin{figure}[ht]
		\includegraphics[width=\linewidth]{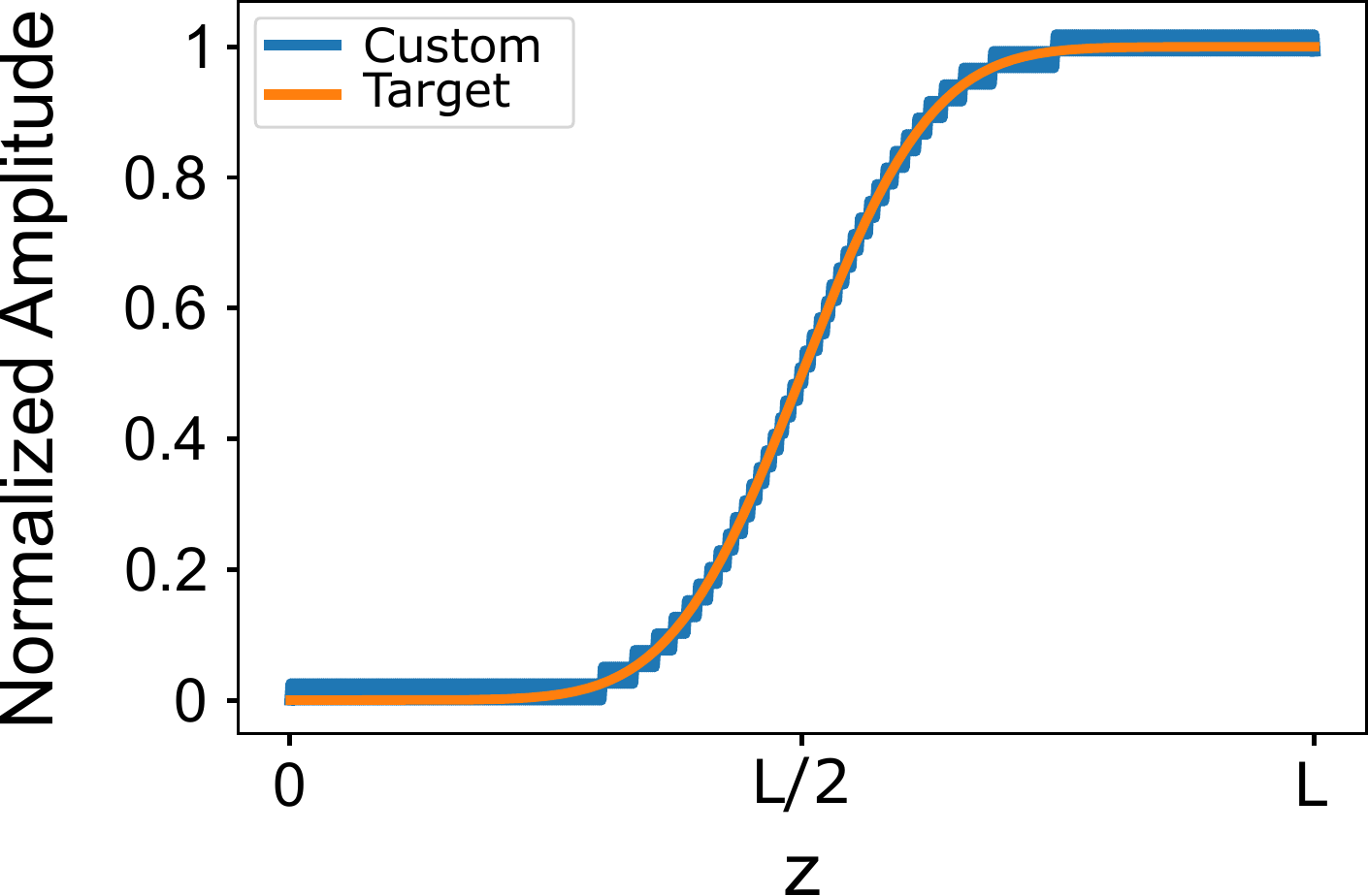}
		\caption{Normalized field amplitudes for both the custom poling $g(z)$ used in the main and those of $g_{\text{Th}}(z)$ which theoretically gives rise to a Gaussian phase-matching function. The custom poling field amplitudes match the target field amplitudes well.}
		\label{fig:amplitude}
	\end{figure}

	\section{Poling function}\label{app:poling}
	As mentioned in the main text of the paper, we obtain an optimal $g(z)$ by following the methods outlined in Ref.~~\onlinecite{tambasco2016domain,Graffitti_2017,Aggie2021domain}. Since we break our region into $N_{z}=1000$ domains, plotting $g(z)$ directly is unhelpful for reproducing results. We can, however, plot the field amplitudes produced by $g(z)$ and show that they correspond to the target expectations. To obtain a Gaussian phase-matching function, one requires $g_{\text{Th}}(z)=\exp\left[-(z-L/2)^{2}/(2\sigma_{\text{Th}})    \right]$ where $L$ is the length of the nonlinear region and $\sigma_{\text{Th}}$ is some spread. By demanding that the JSA be separable in the low-gain regime, we find that
	\begin{align}
		\sigma_{\text{Th}} = \frac{2}{\sigma}\left(\frac{1}{v_{S}}-\frac{1}{v_{P}}\right)^{-1},
	\end{align}
	where $\sigma$, $v_{S}$, and $v_{P}$ are defined in the main text. The theoretical poling function $g_{\text{Th}}(z)$ has field amplitudes given by
	\begin{align}
		A_{\text{Target}}(z) &\propto \int_{0}^{z}dz' g_{\text{Th}}(z') \nonumber\\
		&\propto \text{Erf}\left(  \frac{L-2z}{2\sqrt{2}\sigma_{\text{Th}}} \right)-\text{Erf}\left(  \frac{L}{2\sqrt{2}\sigma_{\text{Th}}} \right),
	\end{align}
	where $\text{Erf}(x)$ is the error function. We can then easily calculate the field amplitudes generated by our custom poling function $g(z)$
	\begin{align}
		A_{\text{Custom}}(z) &\propto \int_{0}^{z}dz' g(z').
	\end{align}
	Fig.~\ref{fig:amplitude} shows the normalized field amplitude as a function of length over the whole nonlinear region. The custom field amplitudes match the target amplitudes quite well.

\nocite{*}
\bibliography{main}

\end{document}